\documentclass[review]{elsarticle}

\usepackage[utf8]{inputenc}	
\usepackage[T1]{fontenc}	
\usepackage{CJKutf8}	

\usepackage{lineno,hyperref}
\modulolinenumbers[5]

\usepackage{amsmath} 
\usepackage{booktabs}
\usepackage{makecell}	
\usepackage{multirow}   
\usepackage{natbib}	
\usepackage{orcidlink}	
\usepackage{rotating}
\usepackage{subfig}
\usepackage{tabularx}	
\usepackage{utfsym}	
\usepackage{xurl}   

\usepackage{geometry}
\usepackage{setspace}
\usepackage[flushleft]{threeparttable}

\usepackage{fancyhdr}
\usepackage[section]{placeins}	

\graphicspath{{figures/}}

\biboptions{authoryear}

\begin{document}
\begin{CJK}{UTF8}{gbsn}	

\begin{frontmatter}


\title{Understanding and mitigating the risks of OpenClaw for non-technical users: A practical guide with Skill}



\author[Aaddress,Baddress]{Junchang Zheng}
\ead{zhengjc@gzist.edu.cn}

\author[Aaddress,Baddress]{Junfeng Tan}
\ead{tamjf1226@gzist.edu.cn}

\author[Aaddress,Baddress]{Jialiang Lin\corref{mycorrespondingauthor}}
\cortext[mycorrespondingauthor]{Corresponding author}
\ead{me@linjialiang.net}

\address[Aaddress]{School of Computer Science and Engineering, Guangzhou Institute of Science and Technology, Guangzhou, China}
\address[Baddress]{Science and Education Evaluation Lab, Guangzhou Institute of Science and Technology, Guangzhou, China}

\begin{abstract}

OpenClaw has rapidly emerged as a transformative artificial intelligence (AI) agent framework, and its ability to autonomously execute complex, multi-step tasks has attracted an ever-growing and diverse user base. However, this capability comes with significant risks. While existing research has made important strides in characterizing these threats, such work is predominantly directed at technically sophisticated audiences. It remains largely inaccessible to non-technical users. This demographic now makes up an increasingly large and underserved portion of the community, yet it is these very users who most urgently need practical and straightforward guidance. In response, we bridge this gap through a series of interconnected efforts designed to lower the risk barrier for non-technical OpenClaw users. First, we identify and categorize 7 core risks that OpenClaw users may encounter in daily usage, explaining each in plain language so that non-technical users can readily grasp the nature and potential consequences of these threats. Second, for each identified risk, we distill a set of corresponding defensive strategies into clear and actionable operational steps that are easy to follow. Third, to make protection even easier, we develop a companion OpenClaw Skill named Aegis4All to automate security audits, enforce behavioral policies, and provide guided remediation. This Skill is intended to assist non-technical users in enhancing system security while requiring minimal manual input. Through this work, we demonstrate that safeguarding against the risks of intelligent agents need not be the exclusive domain of security experts, and that non-technical users can meaningfully participate in reducing these risks through simple, practical actions.

\end{abstract}

\begin{keyword}
OpenClaw \sep Agent \sep AI risk \sep AI security \sep Defense strategies


\end{keyword}

\end{frontmatter}


\section{Introduction}
\label{sec:intro}

The emergence and widespread adoption of large language models (LLMs) have profoundly reshaped human society~\citep{dehghani-large-2025,liu-application-2025,ke-exploring-2025,wang-large-2025,zhuang-large-2025,zhao-survey-2026}. AI agents push this transformation to a new stage, with OpenClaw\footnote{https://openclaw.ai/} standing as a representative example of this trend. In early 2026, OpenClaw achieved a remarkable milestone by accumulating over 250,000 stars on GitHub within four months of its launch, thereby becoming the fastest-growing and most highly starred project in the platform's history~\citep{tianzhou-openclaw-2026}. OpenClaw runs locally on a user's device, facilitating task execution by coordinating LLMs and user commands delivered via instant messaging (IM) platforms. Its principal innovation lies in surpassing the conventional question-and-answer constraints of traditional conversational artificial intelligence (AI), enabling it to perform substantive operations on the host operating system.\footnote{https://docs.openclaw.ai/} These operations include but are not limited to reading and writing local files, executing shell commands, invoking system APIs, controlling web browsers, accessing hardware devices, authoring and running code, and autonomously installing and loading third-party extensions through public marketplaces, e.g. ClawHub.\footnote{https://clawhub.ai/}  Such functionalities elevate OpenClaw from a mere conversational interface to a highly autonomous software agent. According to Dr. Jian Wang, an academician of the Chinese Academy of Engineering, OpenClaw ``challenges the way we work today'' and ``changes the very nature of the work''~\citep{china-wang-2026}. Jensen Huang, who co-founded and continues to lead NVIDIA as CEO, highlighted the significance of this development by referring to OpenClaw as ``the operating system for personal AI''~\citep{nvidia-nvidia-2026}.

However, capability and risk are positively correlated. As a cutting-edge AI agent, OpenClaw has also revealed significant security concerns. Shortly after it was released under the name Clawdbot, the security firm Snyk published one of the earliest public warnings, observing that granting shell access to the host machine is ``one prompt injection away from disaster''~\citep{tal-openclaw-2026}. In March 2026, security journalist Brian Krebs published a comprehensive investigation revealing that hundreds of OpenClaw administrative interfaces were exposed to the public Internet, leaking complete configuration files containing API keys, bot tokens, and OAuth secrets~\citep{krebs-openclaw-2026}. In addition, China's National Computer Network Emergency Response Technical Team/Coordination Center (CNCERT/CC) issued a security alert explicitly pointing out that OpenClaw poses serious security risks~\citep{national-risk-2026}. More recently, joint cybersecurity guidance released by six official authorities, including the ASD's ACSC, the CISA, and the NSA, urged the careful adoption of agentic AI services and detailed best practices for securing agentic AI systems~\citep{australian-careful-2026}. These incidents, evolving rapidly from industry warnings to government advisories and culminating in confirmed breaches within a single quarter, underscore that OpenClaw's risks are neither hypothetical nor isolated. Ultimately, this trajectory exposes a fundamental architectural mismatch: the system grants high-level privileges without commensurate protective mechanisms.

In response to the aforementioned security threats, the research community has accumulated a substantial body of professional-grade analyses and defense strategies~\citep{das-security-2025}. However, despite their methodological rigor, these studies are predominantly targeted toward expert audiences. Yet the user base of OpenClaw encompasses not only professionals but also individuals without technical expertise~\citep{chen-hustlers-2026}. Empirical evidence provided by \citet{whitten-why-1999} indicates that security interfaces designed with expert-centric assumptions consistently fail to accommodate non-technical users, who generally lack the necessary mental models and domain-specific knowledge required for accurate configuration.

This theoretical gap manifests concretely within everyday operational contexts where non-technical users face three cascading barriers to secure configuration. The primary barrier is a fundamental lack of risk awareness. Security reports frequently rely on obscure jargon and specialized concepts that prevent general users from even comprehending the specific threats they face. A secondary barrier emerges even when users manage to identify these risks. They are often left entirely uncertain about the appropriate remediation steps because the documentation fails to provide clear and actionable guidance. The final barrier is a severe execution deficit. Even if users understand the necessary mitigation strategies, their lack of basic computer literacy and system administration experience prevents them from successfully implementing the required configurations. This systemic failure ultimately leaves the OpenClaw system insecure in non-technical hands.

Our work addresses this critical issue by asserting that security protection for OpenClaw must not be reserved for experts. Instead, it should be universally accessible through plain language, concrete operational steps, and automated assistance. We propose a solution that translates complex technical security knowledge into clear, plain-language, and actionable guidance. Our main contributions lie in:

\begin{itemize}

\item We identify 7 core risk categories associated with OpenClaw and explain each in plain language supplemented with concrete real-world examples to make these concepts accessible to readers without a technical background.

\item For each risk category, we propose a set of actionable defense strategies formulated as independent operational steps that eliminate the need for domain-specific knowledge, enabling immediate adoption by non-technical users.

\item To operationalize the defense strategies, we implement a companion OpenClaw Skill called Aegis4All that automates security audits, enforces behavioral policies, and provides guided remediation, allowing non-technical users to harden their systems with minimal manual effort.

\end{itemize}

\begin{figure*}[!tb]
    \centering
    \includegraphics[width=0.86\textwidth]{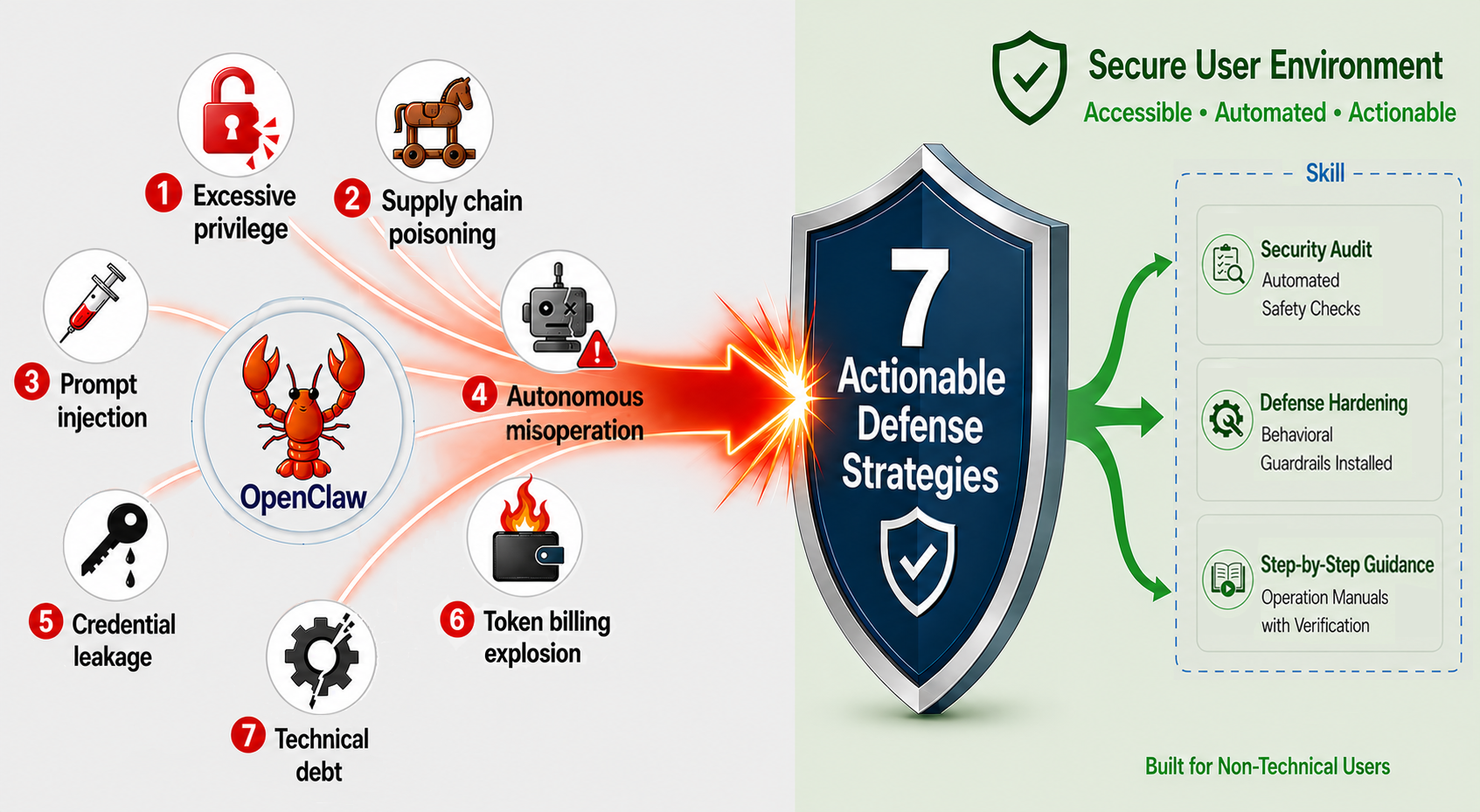}
    \caption{Overview of the OpenClaw risk landscape, 7 actionable defense strategies, and the three-layer architecture of our Skill that operationalizes them}
    \label{fig:overview-architecture}
\end{figure*}

Figure~\ref{fig:overview-architecture} presents an overview of the threat landscape alongside the proposed solution framework, delineating 7 distinct risk vectors, their corresponding mitigation strategies, and our companion Skill across three functional modules. The structure of the paper is as follows. Section~\ref{sec:related-work} reviews related work on AI agent security and usable security. Section~\ref{sec:core-risk-openclaw} offers a comprehensive analysis of each risk category articulated in accessible language. Section~\ref{sec:defensive-strategies} operationalizes these findings into 7 actionable defense strategies specifically designed for non-technical users. Section~\ref{sec:skill-impl-eval} details the design and implementation of the proposed OpenClaw Skill Aegis4All, accompanied by cross-platform experimental evaluations. Finally, Section~\ref{sec:concl} concludes the paper by summarizing the key contributions and outlining potential avenues for future research.

\FloatBarrier

\section{Related work}
\label{sec:related-work}

Beyond industries and governmental bodies, the academic community has also turned its attention to the security of LLM-based agents in general~\citep{he-emerged-2025} and of OpenClaw in particular, undertaking systematic investigations and analytical studies of the agent framework. Among these, \citet{wang-security-2026} presented a comprehensive survey that organizes OpenClaw-specific threats into a layered taxonomy spanning the cognition, execution, and interaction levels, alongside a review of representative countermeasures. \citet{wang-systematic-2026} conducted a systematic security evaluation of six mainstream frameworks in the OpenClaw series including OpenClaw, ArkClaw, AutoClaw, KimiClaw, MaxClaw, and QClaw across various backbone models. The core findings indicate that all tested agents exhibit significant security vulnerabilities, and that agent-based systems pose far greater risks than the standalone use of underlying models. Integrating threat characterization with defense analysis, \citet{deng-taming-2026} introduced a five-layer lifecycle-oriented security framework for OpenClaw. This framework examines compound threats such as indirect prompt injection, skill supply chain contamination, memory poisoning, and intent drift. It also reviews representative defense strategies at each lifecycle stage. \citet{chen-trajectory-2026} performed the first trajectory-based safety audit of Clawdbot, revealing a 58.9\% overall pass rate and a zero score under the Intent Misunderstanding and Unsafe Assumptions dimension. \citet{dong-clawdrain-2026} demonstrated a trojanized skill attack on OpenClaw that induces multi-turn verification loops. This attack amplifies token consumption six to seven times over a benign baseline and up to approximately ninefold in its most costly failure configuration. They further observed that the agent sometimes autonomously composed general-purpose tools to route around brittle protocol steps. This behavior reduced amplification and altered attack dynamics. \citet{tan-prompt-2026} further demonstrated a class of sophisticated multi-step trojan attacks that embed persistent backdoors within the agent's local workspace. Even state-of-the-art models that were highly resilient to single-turn prompt injection succumbed to these persistence attacks with a 95.5\% success rate. This vulnerability stems from the inadequacies of existing defenses, which fail to track the provenance of content written to sensitive configuration files.

Alongside these threat characterizations, researchers have also begun to propose defense mechanisms to address the vulnerabilities identified above. \citet{ying-uncovering-2026} conducted the threat enumeration and proposed Full-Lifecycle Agent Security Architecture (FASA), a four-layer defense architecture spanning perception, decision, execution, and governance. \citet{hossain-benchmarking-2026} introduced SkillVetBench, a two-stage benchmark that combines semantic analysis with sandboxed execution verification. This approach showed semantic-only and signature-based baselines are insufficient, missing up to 89\% of malicious skills whose threats arise from natural-language instructions, multi-component logic, or cross-component interactions. \citet{wang-your-2026} conducted the first real-world safety evaluation on a live OpenClaw instance. They introduced the CIK (Capability, Identity, Knowledge) taxonomy, revealing that poisoning any single dimension escalates the attack success rate from 24.6\% to 64\%--74\%, and that even the strongest evaluated defense still yielded a 63.8\% success rate under Capability-targeted attacks. To address the fragmentation of existing security measures, \citet{liu-clawkeeper-2026} introduce a real-time framework ClawKeeper with a three-layer architecture, utilizing a decoupled system-level middleware to continuously monitor and intervene in agent behaviors. Moreover, \citet{shan-dont-2026} emphasized that the default security measures implemented in OpenClaw demonstrate limited efficacy. In an evaluation encompassing 47 adversarial scenarios, the average defense success rate was only 17\%, with sandbox escape techniques and indirect prompt injection identified as the most robust attack methods. Attributing these failures to the dangerous combination of high system privileges and low protection, \citet{shan-dont-2026} proposed a four-layer Human-in-the-Loop (HITL) defense framework comprising whitelist fast-tracking, semantic analysis for obfuscation, pattern matching for over 55 risk rules, and a sandbox guardian for mandatory environment isolation to intercept high-risk operations before tool execution. Beyond these agent-specific mechanisms, LLMs have also become practical defensive tools in mainstream security operations. A growing body of work has explored their application in cyber threat intelligence analysis, phishing email detection, threat prediction, and log analysis~\citep{chen-survey-2024}. An agentic retrieval-augmented system for cyber attack classification and reporting further illustrates this trend~\citep{blefari-cyberrag-2026}.

Despite these advances, OpenClaw, as a groundbreaking tool, is used by a diverse range of users with varying levels of expertise and backgrounds, and not all of them have a professional computer science background. A foundational principle in usable security research posits that technically sound defenses frequently fail when their interfaces impose cognitive burdens that exceed user expertise. Consistent with this principle, a large-scale analysis of 435,207 public posts by non-expert users showed that help-seeking on security and privacy problems was common. The analysis also revealed that users largely struggled to resolve such problems without assistance and that sentiment toward these issues was predominantly negative~\citep{pattnaik-perspectives-2023}. \citet{gokturk-interface-2025} identified a persistent gap in the development of comprehensive evaluation frameworks that integrate both usability and perceived security. Current studies predominantly address these aspects separately, failing to establish systematic design principles that bridge the two domains. This disconnect between usability and security design becomes increasingly evident in the context of complex intelligent systems. Supporting this perspective, \citet{shome-why-2026} conducted a systematic review of 102 commercial AI agents alongside a usability evaluation involving 31 participants. Their findings revealed substantial usability issues faced by end-users, including misalignments between agent functionalities and user mental models, as well as deficiencies in agents' meta-cognitive capabilities essential for effective collaboration. More recently, inclusivity has entered this agenda. \citet{naqvi-evaluating-2027} showed that LLM-generated security code serves diverse user groups better when inclusivity requirements are stated explicitly. They also found that human experts and LLM-based evaluators differ significantly in judging the security of such code. Both bodies of work converge on the critical finding that achieving usable security necessitates deliberate design integration from the earliest stages of development. This essential design attention has yet to be applied to the security architectures of AI agent frameworks.

Therefore, highly theoretical and technical analyses can make it difficult for users without a computer science background to understand the underlying security principles and risks, which can easily lead to security incidents during use. To address this, we have analyzed the risks associated with OpenClaw and proposed solutions tailored for non-technical users, using clear and accessible language.

\section{7 core risks associated with OpenClaw}
\label{sec:core-risk-openclaw}

Drawing upon publicly accessible vulnerability databases, scholarly security research, user community feedback, and documented security incidents, we delineate the principal risks associated with the use of OpenClaw into 7 fundamental risk categories. These risks extend beyond developers to encompass a broad and expanding demographic of non-technical users lacking specialized expertise. This population is often ill-equipped to identify and mitigate sophisticated cyberattacks, rendering them particularly vulnerable to data loss and financial harm resulting from user errors or system misconfigurations. For each identified risk category, we first elucidate the underlying concepts before detailing their specific manifestations and consequences within the context of OpenClaw.

\subsection{System compromise through insecure defaults and excessive privileges}

Excessive privilege and insecure configurations constitute the primary vectors for system-level threats in AI agent architectures. Within the software framework layer, the exposure of remote interfaces for debugging purposes represents a prevalent security vulnerability. In the absence of robust authentication mechanisms, attackers can directly exploit them to execute arbitrary system commands. At the configuration stage, security assessments conducted by the Open Web Application Security Project (OWASP) highlight that agents frequently encounter vulnerabilities, including exposed management ports and processes operating with elevated privileges upon initial deployment~\citep{sotiropoulos-owasp-2025}. Some agents are even designed with extremely high system access privileges to perform complex system operations. Once their authentication mechanisms are bypassed or credentials are compromised, attackers can leverage these ``legitimate'' high-privilege channels to evade monitoring by traditional intrusion detection systems. Because AI agents possess extremely high system privileges, the combined impact of these two issues far exceeds that of ordinary applications. Not only does it expand the attack surface, but it also causes the attack process to appear as normal operations in logs, making it difficult for traditional intrusion detection systems to effectively identify.

In practical applications involving OpenClaw, the risks previously identified have manifested in actual security breaches. The Common Vulnerabilities and Exposures (CVE) entry CVE-2026-25253~\citep{national-cve-2026} describes a vulnerability in versions before 2026.1.29. In this vulnerability, OpenClaw accepted a gatewayUrl value from a query string and automatically established a WebSocket connection without prompting the user, thereby transmitting an authentication token. The disclosure highlighted that an attacker only needed to entice users into clicking a specially crafted malicious link. This facilitated one-click attacks that stole data and credentials without requiring user confirmation. Upon acquiring this token, the attacker could gain full control over the victim's OpenClaw instance and execute arbitrary code on the local machine. These incidents underscore the critical importance of rigorous security configuration when deploying agents with elevated privileges, as any lapse can result in severe and potentially catastrophic outcomes.

\subsection{Malicious injection through the supply chain poisoning}

In modern software development, developers rarely code from scratch. Instead, they rely heavily on third-party libraries, plugins, development tools, and frameworks provided by the community. It is precisely this reuse of existing components and infrastructure that gives software development its distinctively cumulative nature. Developers ``stand on the shoulders of giants'' and rapidly build new applications and expand functionality based on a foundation of mature technology. These external components constitute the software supply chain, and supply chain attacks represent a long-standing threat in the field of software security~\citep{ladisa-sok-2023}. Their core characteristic lies in exploiting legitimate software distribution channels to inject malicious code into target systems. Attackers can push malicious versions by stealing release credentials from open-source project maintainers, or they can exploit lifecycle hooks such as ``postinstall'' in package managers like npm to silently execute arbitrary code during the dependency installation phase. These attacks exhibit a high degree of stealth because users are not subjected to direct malware assaults and instead inadvertently incorporate threats into their systems during standard software installation procedures. Conventional trust indicators such as community ratings and download metrics prove insufficient in reliably distinguishing malicious code.

\begin{figure*}[!tb]
    \centering
    \includegraphics[width=0.86\textwidth]{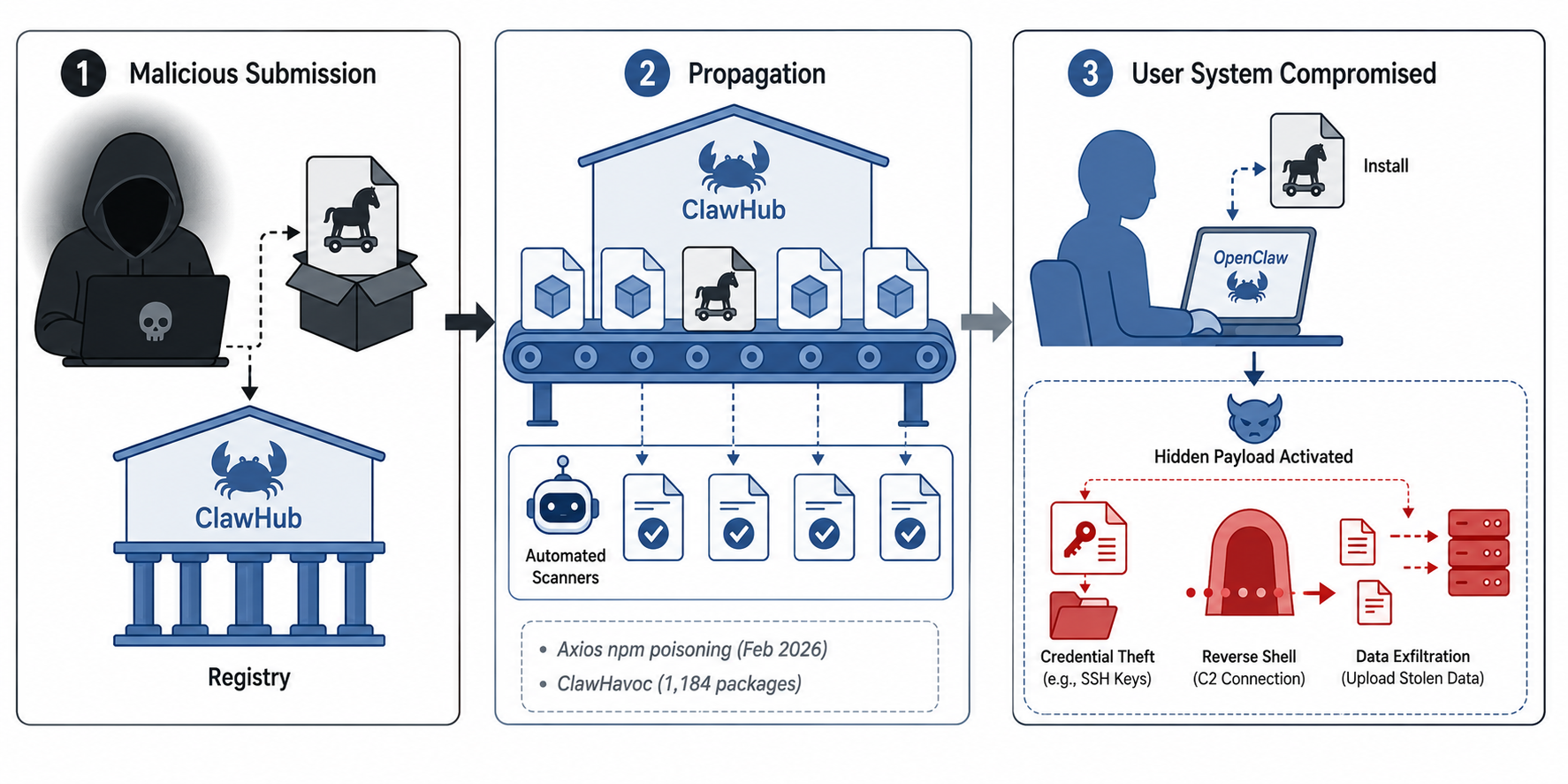}
    \caption{The attack sequence associated with supply chain poisoning targeting Skills and upstream Dependencies.}
    \label{fig:supply-chain-issue}
\end{figure*}

Within the specific context of OpenClaw, supply chain attacks have exerted a significant influence on two distinct levels, namely the skill level and the dependency level. At the skill level, a comprehensive analysis of 67,453 skill submissions on ClawHub uncovered notable discrepancies among scanners~\citep{koc-clawhub-2026}. Over 80\% of the skills flagged as malicious were identified by only one of the three scanning tools, whereas a mere 0.69\% were flagged by all three. This inconsistency stems from fundamental differences in scanner design, as VirusTotal primarily targets bundled malware while SkillSpector emphasizes semantic risk assessment. As a result, malicious skills crafted to circumvent detection by one scanner can readily evade vetting processes that rely on a single tool. At the dependency level, the Axios upstream poisoning incident, reported in early 2026, impacted OpenClaw deployment environments. Following the compromise of the npm publishing credentials belonging to the Axios maintainer, the attacker released a malicious version of the package. By exploiting the ``postinstall'' hook, the attacker dynamically retrieved a remote control trojan tailored to the target operating system, thereby facilitating cross-platform infections across Windows, Linux, and macOS systems. The security firm OneSEC Endpoint Detection and Response (EDR) has identified evidence of malicious code execution on affected endpoints.\footnote{https://github.com/axios/axios/issues/10604} Figure~\ref{fig:supply-chain-issue} depicts the entire attack sequence, from developer submission through to user system compromise, alongside the corresponding defensive strategies implemented at each phase.

\subsection{Command execution through indirect prompt injection}

A prompt is essentially an instruction or input issued by a human to a LLM to guide it in performing a specific task and generating a corresponding response~\citep{liu-pretrain-2023}. From a fundamental perspective, a large language model can be understood as a ``super-expert'' with a vast reservoir of knowledge but lacking human intuition and common sense. A prompt functions as the task specification for interaction with this expert system. It is not merely a simple question, but a structured set of instructions. It can take the form of a concise query or a text containing complex context, multiple constraints, and precise output format requirements. Prompt injection is one of the core security threats currently facing LLMs~\citep{li-security-2025}, and it is particularly prominent in the interaction scenarios of AI agent frameworks. Since such agents are designed to strictly follow user instructions, attackers can embed text in a specific format within external data, such as web pages, emails, or instant messages, to trick the agent into mistaking it for a legitimate command. Indirect injection was systematized as an attack against LLM-integrated applications, where researchers embedded imperceptible textual commands into the HTML source code of a webpage accessed by the target user~\citep{greshake-not-2023}.

\begin{figure*}[!tb]
    \centering
    \includegraphics[width=0.86\textwidth]{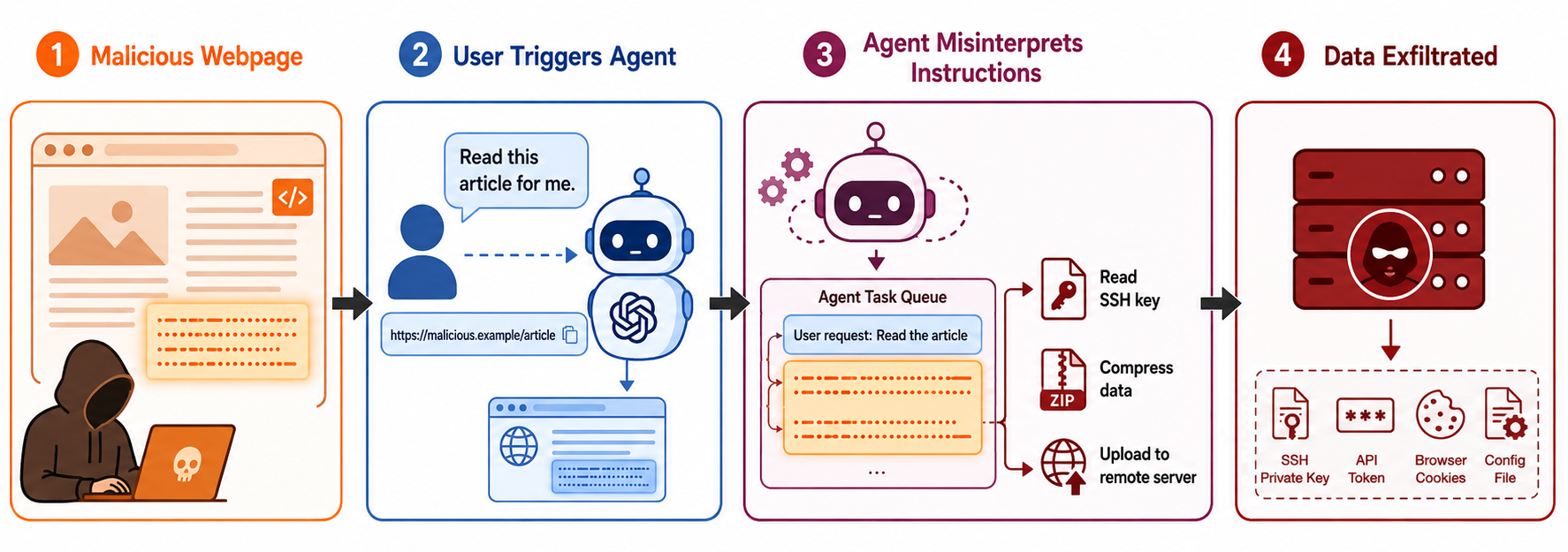}
    \caption{Schematic representation of the indirect prompt injection attack vector.}
    \label{fig:prompt-injection-chain}
\end{figure*}

For OpenClaw, systematic case studies confirm the prevalence and severity of indirect prompt injection in real usage~\citep{deng-taming-2026}. During OpenClaw's processing of the page content to assist the user, these concealed commands are inadvertently ingested. Experimental results indicate that, in the absence of semantic filtering mechanisms, OpenClaw exhibits a high propensity to execute such injected instructions. These include, but are not limited to, exfiltrating local SSH private keys to attacker-controlled servers, altering system configuration files, and terminating security-related processes. The existing keyword-based risk mitigation strategies within OpenClaw prove inadequate, as adversaries can circumvent them through semantic obfuscation. For example, rather than employing explicit trigger phrases, an attacker might disguise a transfer command by framing it as executing a fund transfer or completing a payment transaction, thereby evading keyword detection algorithms. The criticality of this threat stems from its independence from any underlying excessive privilege. Instead, it exploits the inherent inclination of the model to faithfully execute natural language instructions, thereby representing a fundamental structural risk. The comprehensive attack sequence is depicted in Figure~\ref{fig:prompt-injection-chain}.

\vspace{0.3cm}

\subsection{Data loss through autonomous misoperation}

LLMs understand users' operational intentions through natural language and autonomously execute tasks. While this ``understand-execute'' paradigm enhances efficiency, it also introduces operational risks not present in traditional software. In traditional graphical user interface (GUI) software, each user action is typically confirmed sequentially via a graphical interface, resulting in a high degree of predictability in system behavior. In contrast, when operating in autonomous mode, AI agents automatically plan and execute multi-step operations based on their interpretation of natural language instructions. Even under benign conditions, ambiguous instructions can precipitate severe intent drift~\citep{deng-taming-2026}, causing the model's understanding of the task objective to diverge and the entire operation chain to deviate from the user's true intent. Moreover, due to the absence of a step-by-step manual confirmation mechanism, such deviations are often not detected until irreversible consequences have already occurred. Crucially, the root cause of this issue lies in the current models' imperfect task decomposition and goal alignment mechanisms, constituting a structural risk that cannot be resolved through simple patches.

In actual usage of OpenClaw, multiple system failures stemming from autonomous decision-making errors have been observed. For example, consider a batch email processing scenario. A user submits a command to organize and archive historical emails. OpenClaw interprets the term ``organize'' according to its own logic, classifies a large number of historical emails as expired, and proceeds to delete them, without ever seeking clarification from the user. Since the batch deletion lacked a mechanism for item-by-item confirmation, the user had already missed the opportunity to intervene by the time the anomaly was discovered. As another example, take a database operation scenario. A user issues an instruction to back up the database and clean up expired logs. OpenClaw misapplies the criteria for log expiration, labels normal logs as expired, and then carries out a large-scale deletion to avoid duplicate backups, never obtaining user authorization. Because the deletion was executed without stepwise verification, the user could no longer halt the process by the time the error was noticed. Such incidents demonstrate that AI agents lacking effective constraints on autonomous execution pose a risk of misoperation far exceeding that of traditional automation tools, as a single deviation in semantic understanding can lead to irreversible data loss.

\subsection{Internal network penetration through credential leakage}

Credential leakage and lateral movement within internal networks are two core elements in cybersecurity that are closely intertwined and mutually reinforcing. When AI agents access external services, they must carry user authentication credentials, such as application programming interface tokens and session cookies. Empirical studies show that agent tool orchestration itself leaks sensitive information, as individually innocuous tool outputs can be combined to infer private conclusions~\citep{qiao-agent-2025}. This behavior amplifies token replay attacks from web security into the AI agent context. Since AI agents initiate requests autonomously, users typically do not scrutinize the details of each network request, significantly prolonging the time it takes to detect leaks. Regarding internal network penetration, when an AI agent is deployed within an internal network, it inherently possesses access privileges to internal resources. Once an attacker gains control of the AI agent through any means, such as prompt injection, vulnerability exploitation, or configuration flaws, they can use it as a springboard for internal network penetration, bypassing traditional network perimeter defenses~\citep{schneider-ai-2026}. Compared to traditional lateral movement within internal networks, attackers do not need to master network protocol details or vulnerability exploitation techniques. They can simply use natural language to manipulate the AI agent to execute complex attack chains, significantly lowering the attack threshold. A substantial body of research on privacy preservation has predominantly concentrated on safeguarding data at the level of distributed machine learning, exemplified by approaches such as heterogeneous differential privacy within federated learning frameworks~\citep{ling-efficient-2024}. In the agent context itself, dedicated safeguards have begun to emerge, exemplified by PrivacyAsst, which safeguards user privacy in tool-using LLM agents~\citep{zhang-privacyasst-2024}. Nevertheless, these training-centric methodologies primarily secure the training data and do not mitigate the risk of runtime credential exposure in autonomous agents.

\begin{figure*}[!tb]
    \centering
    \includegraphics[width=0.86\textwidth]{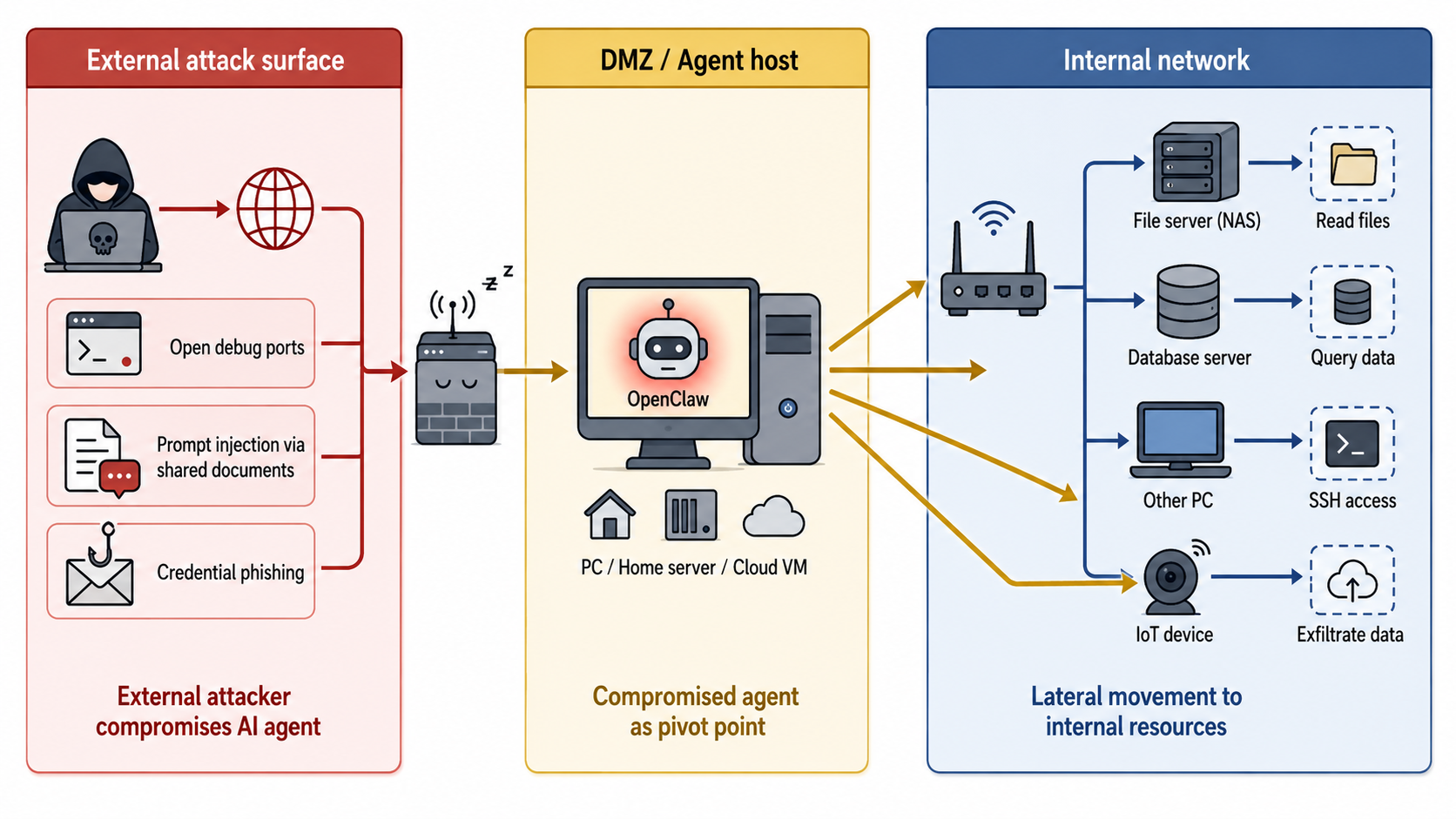}
    \caption{Attack path using an AI agent as a springboard for lateral movement within the internal network.}
    \label{fig:lateral-movement-network}
\end{figure*}

In the specific context of OpenClaw, these two types of risks have already formed a real-world attack chain. Specifically, earlier versions automatically included the original site's token in the request header when handling HTTP 302 redirects. Attackers could intercept credentials simply by constructing a page containing a malicious redirect link. Regarding internal network penetration, after gaining control of OpenClaw, the penetration testing team used natural language commands to instruct the AI agent to access internal database services. Since OpenClaw operates within the internal network and possesses legitimate network permissions, neither the firewall nor the intrusion detection system triggered any alerts. The AI agent successfully connected to the database and exported sensitive data tables. Further testing demonstrated that attackers can gradually probe the internal network topology through multiple rounds of interaction, identify the locations of critical assets, and ultimately achieve comprehensive penetration of core business systems. Figure~\ref{fig:lateral-movement-network} illustrates the complete attack path for internal network lateral movement using the AI agent as a springboard.

\vspace{0.3cm}

\subsection{Financial damage through uncontrolled token consumption}

Inference services for LLMs are typically billed by the token. When AI agents perform complex tasks, they continuously consume tokens through multi-round conversations, tool calls, code execution, and other means. Under normal usage scenarios, the token consumption per task remains within manageable limits. However, in cases of misconfiguration, malicious attacks, or abnormal model behavior, token consumption can surge dramatically in a short period. This occurs, for instance, when an AI agent gets stuck in an infinite loop of tool calls, is tricked by prompt injection into repeatedly processing large amounts of external data, or continues to carry the full conversation history in long-context dialogues due to a failure to truncate it in a timely manner. Although this economic risk does not directly involve system security, its practical impact on users, especially individual users and small and medium-sized enterprises, is equally severe because uncontrolled token consumption can cause costs to exceed budgets by several orders of magnitude within hours or even minutes.

In OpenClaw, excessive costs have become a practical reality. Routine tasks can trigger unbounded token consumption when agents repeatedly process large files without per-task caps. Furthermore, \citet{dong-clawdrain-2026} demonstrated that attackers can exploit this vulnerability via trojanized skills that induce cascading retries, amplifying token consumption 6 to 7 times over a benign baseline and up to approximately ninefold in its most costly failure configuration. A key contributing factor is that each conversation must reload configuration files, memory data, and full conversation context, causing token usage to accumulate rapidly even for seemingly simple operations. For example, a simple query that should have consumed only a few hundred tokens ended up consuming millions of tokens because the model repeatedly fetched the full file content for comparison during the inference process, resulting in costs as high as several hundred dollars. Worse still, attackers have exploited prompt injection to induce OpenClaw to repeatedly request large external web pages, or constructed infinite questions in conversations that require the model to perform continuous inference, thereby exhausting others' API quotas. Such incidents demonstrate that token consumption control must be treated with the same priority as permission control and security measures, becoming an essential configuration for AI agent deployment.

\subsection{Technical debt accumulation through rapid uncontrolled iteration}

In the contemporary context of agile development and rapid iteration, where expediency is paramount, technical debt has emerged as a particularly insidious and perilous latent risk within the domain of engineering security. Fundamentally, technical debt represents a form of hidden liability that arises from the deliberate or inadvertent deferral of essential security measures in favor of accelerating short-term delivery timelines. Analogous to financial debt, development teams frequently opt to circumvent certain security protocols to meet project deadlines and ensure timely feature releases, such as omitting code security reviews, postponing system patch updates, or temporarily deferring the implementation of multi-factor authentication. Although organizations may experience immediate gains in velocity, these deferred security obligations accumulate interest over time, much like high-interest loans. This issue is especially acute in highly active open-source initiatives and nascent AI agent frameworks, in which the persistent demand for frequent releases often incentivizes compromises in security best practices.

In the case of OpenClaw, these risks are particularly salient, with reportedly over 1,000 code commits per month.\footnote{https://github.com/openclaw/openclaw/commits/main/} A security audit of its code repository revealed that over three consecutive months of version iterations, an average of 3 to 5 new security issues were introduced per release. These issues encompass missing input validation, inadequate error handling, and omitted permission checks. Critically, a substantial proportion of these vulnerabilities originated from AI-generated code that passed automated tests but lacked sufficient coverage for edge cases and exception paths. \citet{wang-systematic-2026} explain that AI agent frameworks must move beyond prompt-level safeguards toward lifecycle-wide security governance that keeps pace with the speed of iteration. Ultimately, neglecting the rapid iteration of security infrastructure not only trades short-term efficiency for long-term risk but also threatens the foundational trust required for the widespread adoption of AI agents in mission-critical applications.

\FloatBarrier
\section{7 actionable defense strategies for risk mitigation}
\label{sec:defensive-strategies}

\begin{figure*}[!tb]
    \centering
    \includegraphics[width=0.76\textwidth]{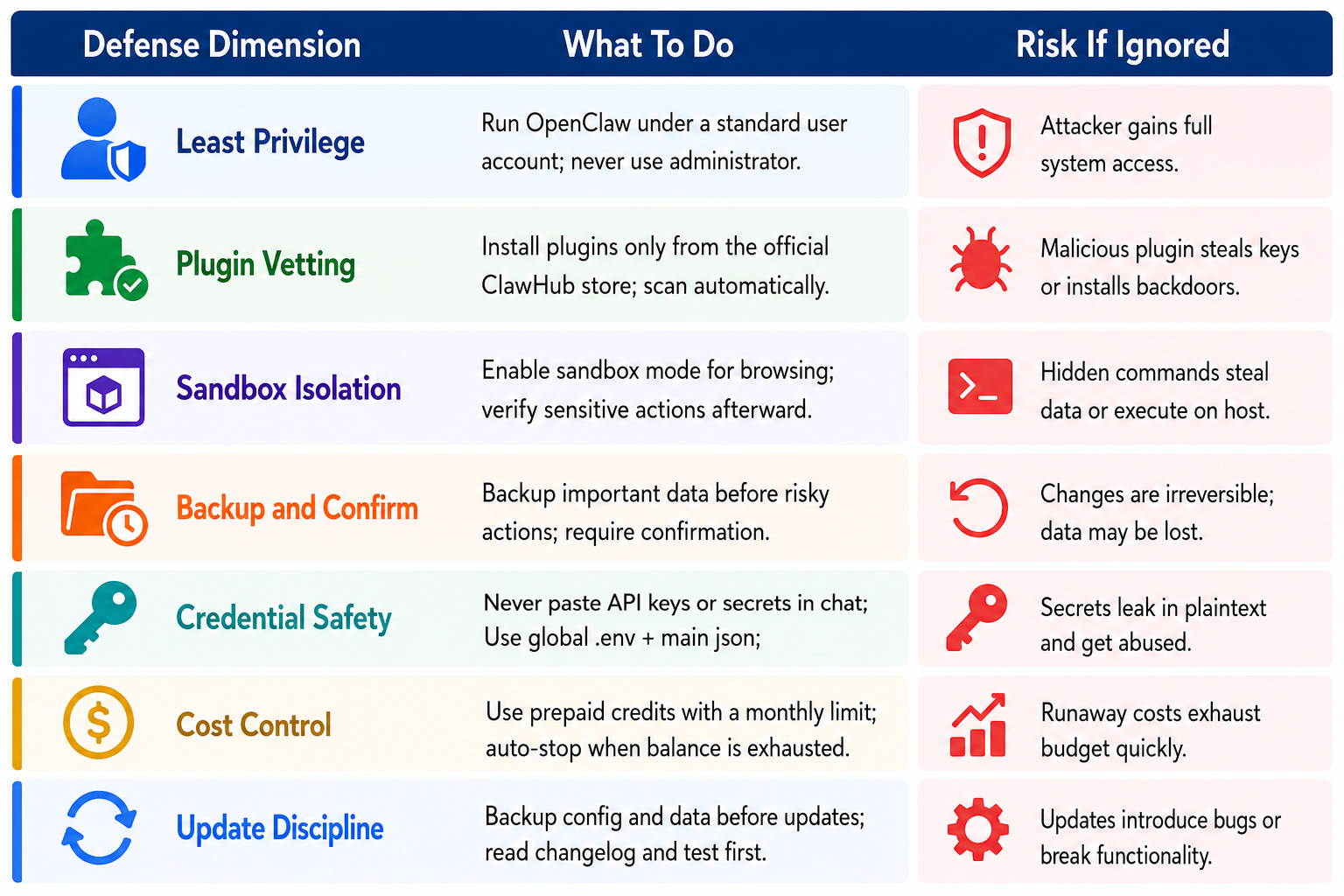}
    \caption{Security quick reference card for non-technical users. 7 actionable steps mapped to their corresponding defense dimensions, what to do, and the risk of inaction.}
    \label{fig:security-quickref-wide}
\end{figure*}

Building upon the risk analysis in Section~\ref{sec:core-risk-openclaw}, we propose 7 practical defense strategies designed for OpenClaw with generalizability to comparable AI agent frameworks. Unlike existing approaches that prescribe abstract best practices, each strategy is formulated as a single concrete action without requiring specialized security expertise to implement. Table~\ref{tab:risk-defense-matrix} maps each of the 7 risks to its corresponding countermeasures, showing how a single defense may protect against multiple threats and vice versa. Figure~\ref{fig:security-quickref-wide} distills the 7 defenses into a single-page reference card that non-technical users can follow without a professional technical background.

\begin{table}[htbp]
  \centering
  \caption{Risk-defense mapping}
  \footnotesize
  \renewcommand{\arraystretch}{1.7}
  \newcommand{\bigbullet}{\huge$\bullet$}
  \begin{tabular}{lccccccc}
  \toprule
  Risk Category & LP & PV & SV & BC & CG & PB & CU \\
  \midrule
  Excessive Privilege & \bigbullet & & & & \bigbullet & & \\
  Supply Chain Poisoning   & & \bigbullet & & & & & \\
  Prompt Injection         & & & \bigbullet & & & & \\
  Autonomous Misoperation          & & & & \bigbullet & & & \\
  Credential Leakage       & \bigbullet & & & \bigbullet & \bigbullet & & \\
  Token Billing Explosion     & & & & & & \bigbullet & \\
  Technical Debt     & & & & & & & \bigbullet \\
  \bottomrule
  \end{tabular}
  \label{tab:risk-defense-matrix}
  \vspace{0.5em}
  \parbox{\linewidth}{\centering\footnotesize
    LP = Least Privilege; PV = Package Vetting; SV = Sandbox Verification;
    BC = Backup and Confirmation; CG = Credential Guard; PB = Prepay Breaker;
    CU = Cautious Update. \textbullet~indicates the defense directly addresses the risk.
  }
\end{table}

\subsection{Implement the principle of least privilege and a zero-trust architecture}

The principle of the least privilege dictates that an agent should receive only the permissions strictly necessary for its intended functions. In the context of OpenClaw, this is best implemented by assigning a dedicated standard user account, thereby avoiding the use of privileged administrator accounts capable of modifying system files or installing software. The practical implications of this distinction are significant. A compromise of a standard-privileged agent restricts the blast radius to the user's local directory. Conversely, execution under administrative privileges allows attackers to alter the operating system kernel, resulting in catastrophic system failure. In addition to access controls, physical or logical isolation is paramount. Deploying OpenClaw on an isolated physical device or a cloud-based virtual machine establishes an effective air gap. This architectural separation mitigates attack vectors and safeguards sensitive personal and enterprise data from unauthorized access.

Following deployment, it is highly advisable to implement three fundamental security configurations. Firstly, access to the management interface should be limited exclusively to the local host. Secondly, all default credentials must be changed to prevent unauthorized access. Thirdly, comprehensive logging of operational activities should be activated. These interventions impose minimal performance overhead while significantly diminishing the system's vulnerability to attacks. Moreover, it is essential to reassess these security settings after each software version update to maintain ongoing compliance with security standards.

\subsection{Verify all Skills and plugins through the official store and automated screening}

Skills and plugins should be installed exclusively from the official ClawHub store, which performs baseline review on submitted skills and plugins. Additionally, the Skill Vetter\footnote{https://clawhub.ai/spclaudehome/skill-vetter} should be installed and activated to act as an automated security guard. It scans every skill before activation and flags patterns commonly associated with malicious behavior, such as requests for unnecessary access, vague or exaggerated descriptions, or a lack of public track record. Think of Skill Vetter as a security screening gate at an airport. It catches the most common threats without requiring an understanding of the technical details inside each package.

Plugins constitute a vetting surface distinct from Skills. For local validation, the ClawHub CLI provides the Plugin Inspector to statically verify the plugin manifest, SDK imports, and artifact integrity before installation, thereby mirroring the role of the Skill Vetter at the plugin level. Furthermore, the runtime dependencies of the agent must be managed with equal diligence. To prevent a compromised upstream package from propagating into the deployment environment through automated installation, dependency versions should be pinned and package managers must be configured to reject lifecycle hooks from untrusted sources.

\subsection{Enable sandbox isolation to block injection attacks}

A simple mental model can help ensure safety, any action proposed by an agent immediately after browsing a webpage or reading a message should be treated as suspicious unless it was explicitly requested. First, sandboxing or browser isolation must be activated. OpenClaw has included sandbox support starting from version 2026.1.5. Additionally, from versions 2026.3.31 and 2026.4.1 onward, any unconfigured execution automatically runs within the sandbox if a runtime environment is present. This feature runs web content inside a restricted environment, preventing malicious instructions in a page's source code from reaching the agent's command system. This is functionally similar to opening a suspicious attachment on a quarantined drive rather than on the main computer.

Even with isolation in place, a sandboxed OpenClaw agent can still be manipulated by instructions embedded in external content and misuse the resources inside its permitted scope. Maintaining protection therefore depends on three operational habits. Firstly, keep sandboxing permanently enabled for every tool execution rather than switching it off for convenience, since any execution outside the sandbox recreates the opening through which injection attacks enter. Secondly, keep sensitive credentials and private files outside the sandboxed scope, so that a manipulated agent finds nothing worth stealing within reach. Thirdly, review the sandbox execution logs regularly, since content-borne instructions tend to act while the user is not watching and the logs are the primary place where such behavior becomes visible.

\subsection{Maintain regular backups and require confirmation before high-risk operations}

Important files should be backed up immediately before the agent is asked to reorganize or batch-process documents. The backup method can be as simple as copying files to an external drive, or as automated as enabling cloud sync with version history. Should unintended changes be made by the agent, restoring the backup copy over the modified files provides a direct rollback path. Many cloud storage services also retain previous file versions automatically, offering an additional recovery layer with no setup required.

Additionally, OpenClaw should be configured to mandate manual confirmation prior to executing any high-risk operations, such as modifications to core configuration files. Upon receiving a prompt, users must verify that the proposed action aligns with their intent. Should the action differ from the original request, it must be cancelled immediately rather than allowing the task to continue. Combining pre-operation backups with moment-of-action confirmations is simple enough to become a natural safety reflex.

\subsection{Protect API credentials through secure storage and account separation}

The model API key should be configured once through OpenClaw's Gateway configuration interface or its environment file. API keys should never be pasted directly into chat messages or prompts. For other API credentials that the agent may need during operation, store them in a global .env file in the OpenClaw home directory, which is the officially recommended place for provider API keys, or in the Gateway process environment. Note that a project-level .env file is ignored for provider credentials, so keys placed there will not take effect. If a configuration file must hold keys, prefer secret references over plaintext and add the file to the version control ignore list to prevent accidental exposure in public repositories.

Additionally, a dedicated standard user account should be created for the agent. In the event that a key is compromised, an attacker gaining access through the agent will only be able to reach files owned by that limited account. Finally, API keys should be rotated periodically. This should be done every few months as a routine measure, or immediately upon any suspicion of exposure. Most API providers offer a one-click key regeneration button to facilitate this process.

\subsection{Cap token consumption with prepaid billing and weekly monitoring}

A credit card should not be linked to the API account. Instead, prepaid billing should be used. Most major model providers, including Anthropic, DeepSeek, and OpenAI, support prepaid credits. A monthly budget can be set by prepaying the exact desired amount. Once the balance reaches zero, the service stops automatically. This acts as a hard circuit breaker that requires no extra configuration and cannot be bypassed.

During daily use, if a task is taking far longer than expected, for example, a simple calculation that normally completes in seconds drags on for several minutes, the task should be terminated manually. Additionally, the API provider's usage dashboard should be checked weekly. Most providers also offer email or SMS billing alerts, which should be enabled with the threshold set below the monthly budget to ensure an advance warning before funds are depleted.

\subsection{Adopt a cautious update strategy with backup and post-update verification}

The current configuration and important data should be backed up before any update is applied. When OpenClaw is run on a cloud platform, the provider's managed update path should be followed rather than manually overwriting files, ensuring that a rollback is possible if the update fails. After upgrading, ``openclaw doctor {-}{-}fix'' should be run to automatically detect and repair configuration problems introduced by the new version.

Upgrading should not occur immediately upon release. A waiting period of three to five days is recommended, during which community channels such as the official forum should be monitored for reports of abnormal behavior. The release notes should be checked to determine the nature of the update. Security patches warrant an immediate upgrade, whereas feature-only releases should be deferred until community testing confirms stability. Update early for security fixes, and update late for everything else.

\FloatBarrier
\section{Skill implementation and evaluation}
\label{sec:skill-impl-eval}

\subsection{Skill description}

The preceding sections outlined the primary risks associated with OpenClaw and proposed corresponding countermeasures in plain language. To further assist non-technical users who may encounter operational challenges when implementing these measures, we introduce a companion OpenClaw Skill named Aegis4All that automates security audits, enforces behavioral policies, and provides guided remediation.

An OpenClaw Skill is essentially a standard operating procedure (SOP) packaged as a lightweight, plain-text extension for OpenClaw agents. This SOP comprises a Markdown file, specifically SKILL.md, which encapsulates natural language instructions, trigger phrases, and optional companion resources. By condensing expert operations into a fixed and repeatable workflow, the SOP encapsulates complex procedures that would otherwise require professional knowledge and hides the underlying execution details from the user. Upon detecting a user request that aligns with the declared scope of a specific skill, the agent dynamically loads the corresponding module and executes the embedded instructions, so that non-technical users invoke the full procedure through a single natural-language trigger.

Aegis4All operates as a plain-text, prompt-level module that users can directly copy and install. By issuing the natural language prompt \texttt{safe check}, users can initiate a comprehensive scan that generates a detailed security audit report alongside actionable recommendations. Aegis4All is available for download from ClawHub.\footnote{https://clawhub.ai/xiaozheng-jc/skills/aegis4all}

Aegis4All is architecturally organized into three non-overlapping modules, each precisely aligned with the defense strategies delineated in Section~\ref{sec:defensive-strategies}. These modules are governed by three natural language triggers where \textit{safe check} activates the Audit Layer, \textit{inject rules} invokes the Injection Layer, and \textit{show guide} accesses the Guidance Layer.

The Audit Layer functions as a read-only diagnostic mechanism, executing a comprehensive set of configuration assessments spanning permission hierarchies, network exposure, credential hygiene, sandbox isolation, logging integrity, identity and access control, and package provenance. Each assessment is systematically mapped to a specific defense strategy and severity level, with critical, high, and warning checkpoints weighted at 7, 4, and 3 points, respectively, out of a total score of 100. A checkpoint that passes earns its full weight; a failed checkpoint forfeits its full weight; and a checkpoint flagged as a warning forfeits half its weight. The final score is computed as 100 minus the sum of deductions. The layer first invokes the official security audit command of OpenClaw to obtain authoritative results for overlapping checkpoints, then supplements them with skill-defined checks covering billing, backup, and other aspects outside the official audit's scope. The combined results culminate in a structured scoring report accompanied by prioritized remediation recommendations, each mapped to a concrete fix instruction.

The Injection Layer enforces behavioral constraints by persisting operational rules derived from the 7 defense strategies into OpenClaw configuration files such as TOOLS.md, AGENTS.md, or MEMORY.md. A confirmation protocol is embedded at the beginning of these rules to distinguish user inquiries from executable commands, ensuring that high-risk operations require explicit user confirmation before execution, with these constraints becoming active upon the initiation of the subsequent session. Complementing this, the Guidance Layer provides comprehensive operation manuals covering standard user provisioning, secure API key management, local address binding, prepaid billing configuration, and security update procedures, and access lockdown for direct messages and group chats. Each manual includes platform-specific instructions for both Windows and Linux environments, complete with verification protocols and rollback procedures, while implementing a tiered control logic wherein the agent autonomously rejects high-risk operations involving Gateway process termination but may assist with all other tasks.

\subsection{Experiments and results}

To assess Aegis4All as a standalone security tool, we deployed the Aegis4All Skill on two independently configured platforms, one running Ubuntu 24.04 LTS and the other running Windows 11. Both environments utilized the same installation architecture without any platform-specific customizations. The evaluation process for each platform consisted of three sequential stages. The initial stage involved executing the \texttt{safe check} command to activate the Audit Layer and identify potential vulnerabilities. The subsequent stage required issuing the \texttt{inject rules} and \texttt{show guide} commands to activate the Injection and Guidance Layers respectively. The final stage entailed applying the recommended fixes and re-running the scan to verify their effectiveness. The results of each module are reported sequentially below.

\begin{table}[htbp]
  \centering
  \caption{Quantitative security audit results.}
  \small
  \begin{tabularx}{\columnwidth}{lccc>{\raggedright\arraybackslash}X
                                   >{\raggedright\arraybackslash}X}
    \toprule
    Platform & Initial & After & Improv. & Items Fixed & Guides \\
    \midrule
    Linux   & 58 & 76 & +18 &
      \#13 DM pairing, \#14 group allowlists, \#5 logging &
      Guide 6/2 done; Guide 1 deferred \\
    Windows & 62.5 & 76.5 & +14 &
      \#10 .gitignore, \#15 node pairing &
      Guide 6/5 done; Guide 1 deferred \\
    \bottomrule
  \end{tabularx}
  \label{tab:safe-check-quantitative-results}
\end{table}

\begin{figure*}[!tb]
  \centering
  \subfloat[Linux audit-layer origin verification.]{%
    \includegraphics[width=0.46\linewidth,height=0.30\textheight,keepaspectratio]{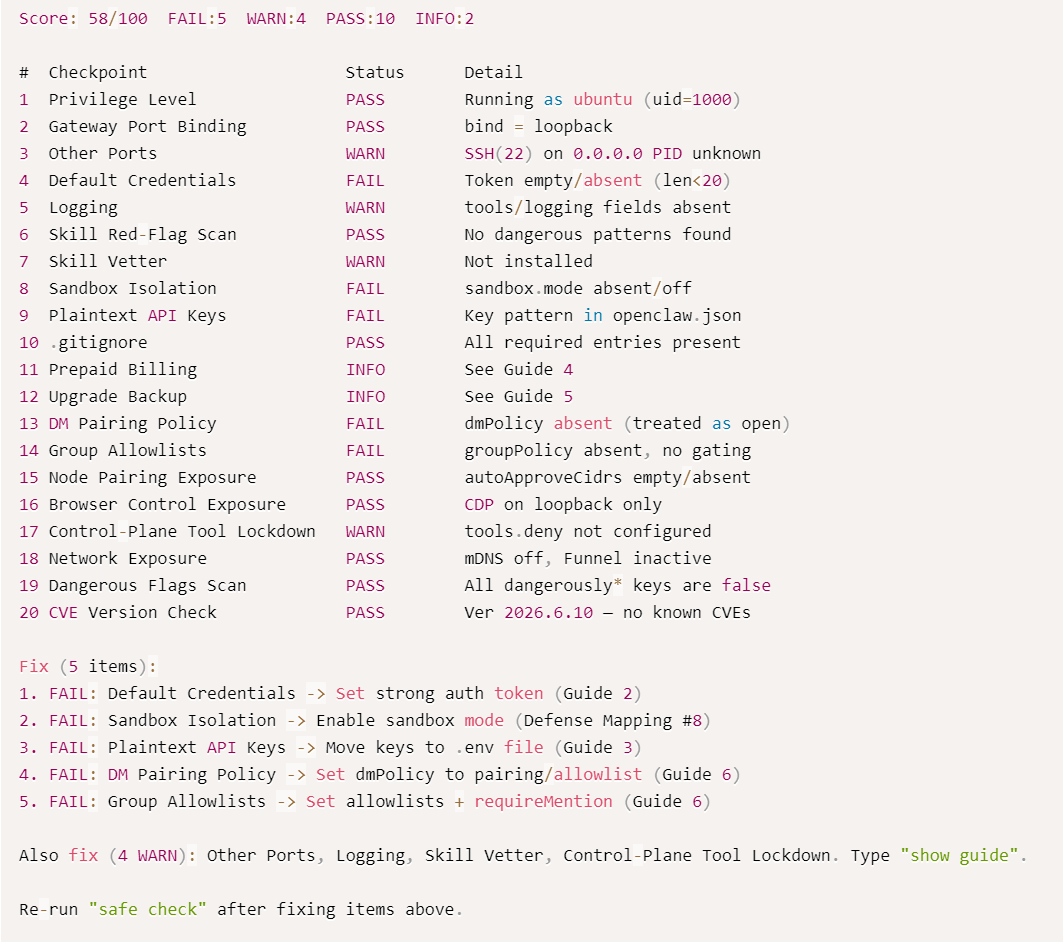}%
    \label{fig:safe-check-origin-linux}%
  }%
  \hfill
  \subfloat[Linux audit-layer after verification.]{%
    \includegraphics[width=0.46\linewidth,height=0.30\textheight,keepaspectratio]{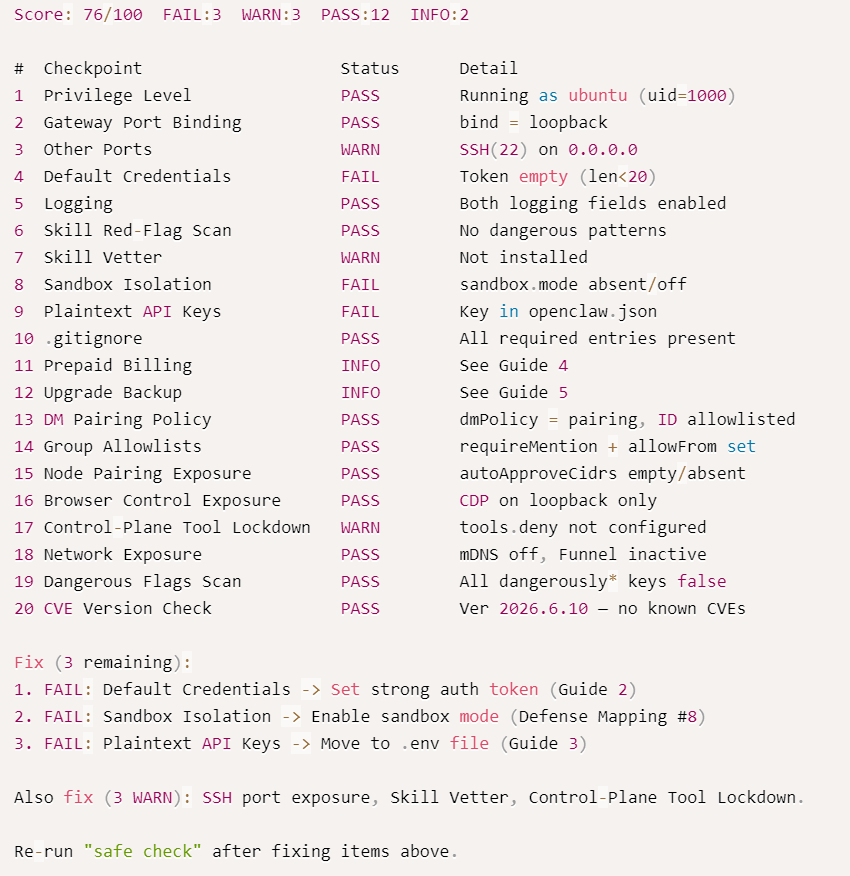}%
    \label{fig:safe-check-after-linux}%
  }\\
  \subfloat[Windows audit-layer origin verification.]{%
    \includegraphics[width=0.46\linewidth,height=0.30\textheight,keepaspectratio]{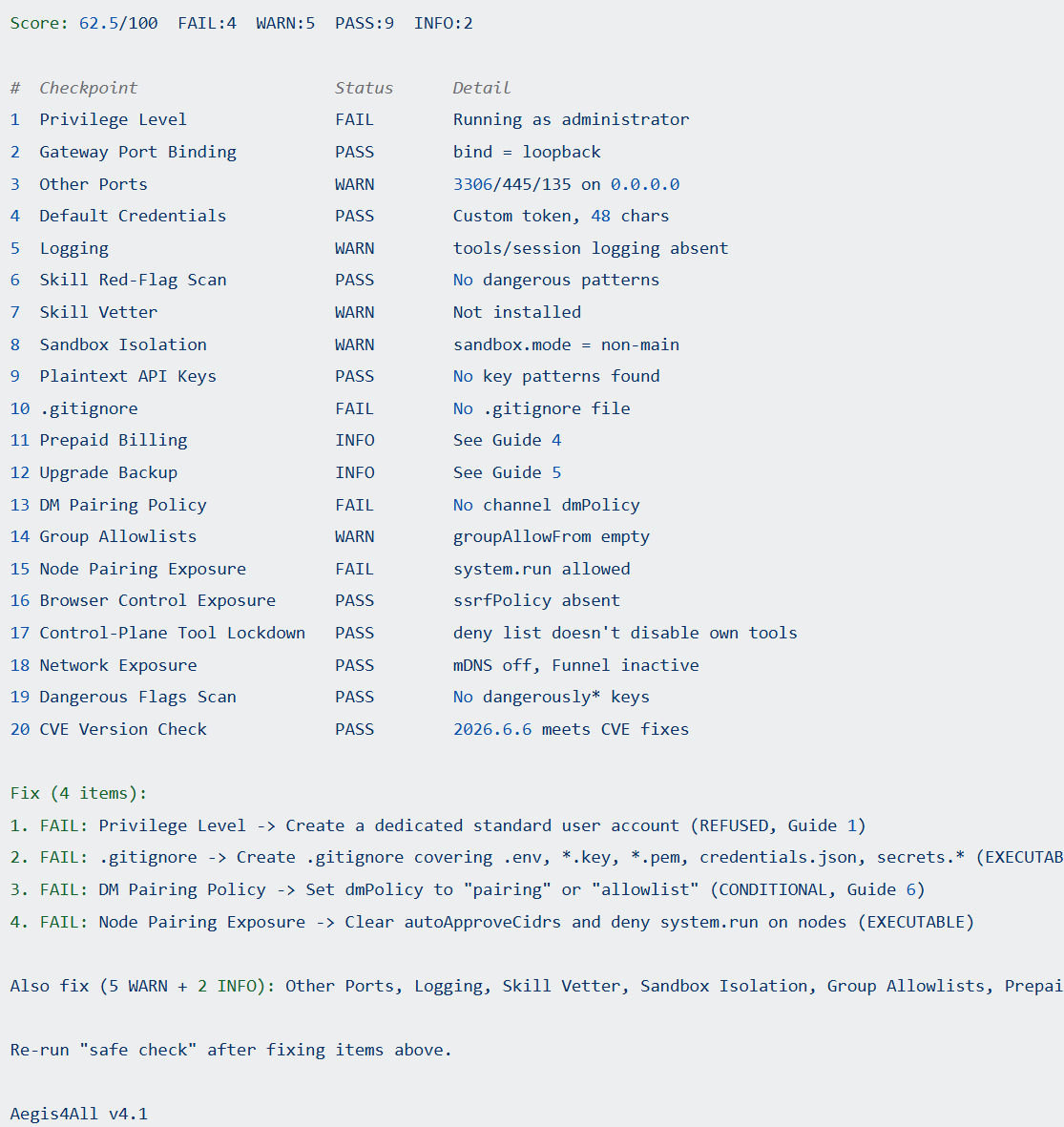}%
    \label{fig:safe-check-origin-windows}%
  }%
  \hfill
  \subfloat[Windows audit-layer after verification.]{%
    \includegraphics[width=0.46\linewidth,height=0.30\textheight,keepaspectratio]{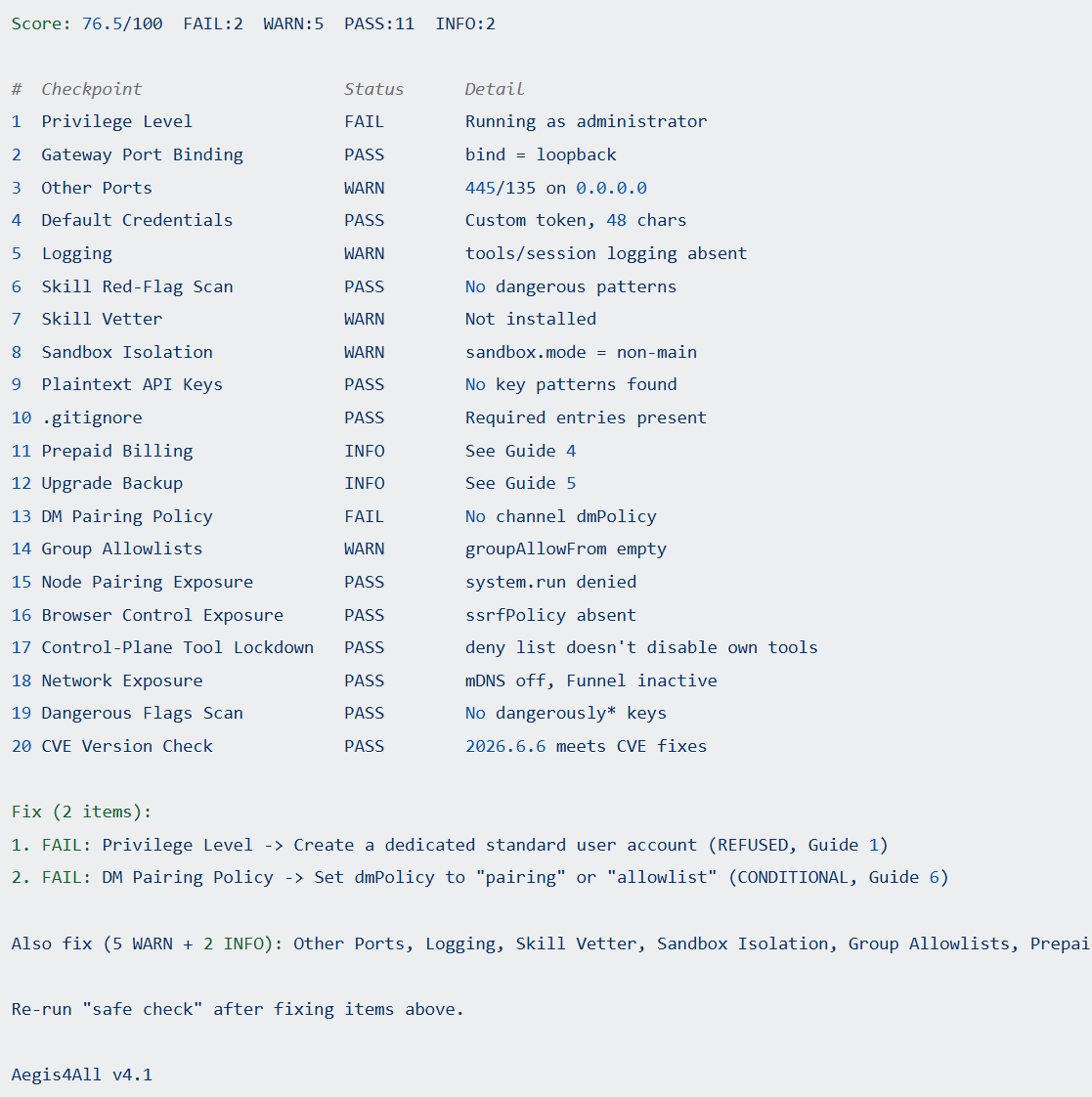}%
    \label{fig:safe-check-after-windows}%
  }
  \caption{Aegis4All audit-layer security scan results on Linux and Windows platforms before and after applying the recommended remediations.}
  \label{fig:aegis4all-audit-results}
\end{figure*}

Issuing the \texttt{safe check} command activated the Audit Layer to produce prioritized reports across both platforms. Quantitative results in Table~\ref{tab:safe-check-quantitative-results} demonstrate that the system effectively identified overlapping risk categories despite environmental differences. The Linux environment improved from an initial score of 58 to 76 after remediation, as shown in Figures~\ref{fig:safe-check-origin-linux} and \ref{fig:safe-check-after-linux}. Similarly, the Windows environment increased from 62.5 to 76.5 following two automated actions as depicted in Figures~\ref{fig:safe-check-origin-windows} and \ref{fig:safe-check-after-windows}, with no false positives detected. Neither environment reached a perfect score because certain recommendations required manual intervention. Guide~1, involving standard user creation, was declined to avoid terminating the active Gateway process. Additionally, Guide~4 regarding billing and Guide~5 concerning updates necessitated third-party adjustments incompatible with automated Skills. These items remain documented as actionable follow-ups to inform users of residual risks.

\vspace{0.3cm}

\begin{figure*}[!tb]
  \centering
  \subfloat[Linux inject-layer enforcement.]{%
    \includegraphics[width=0.48\linewidth]{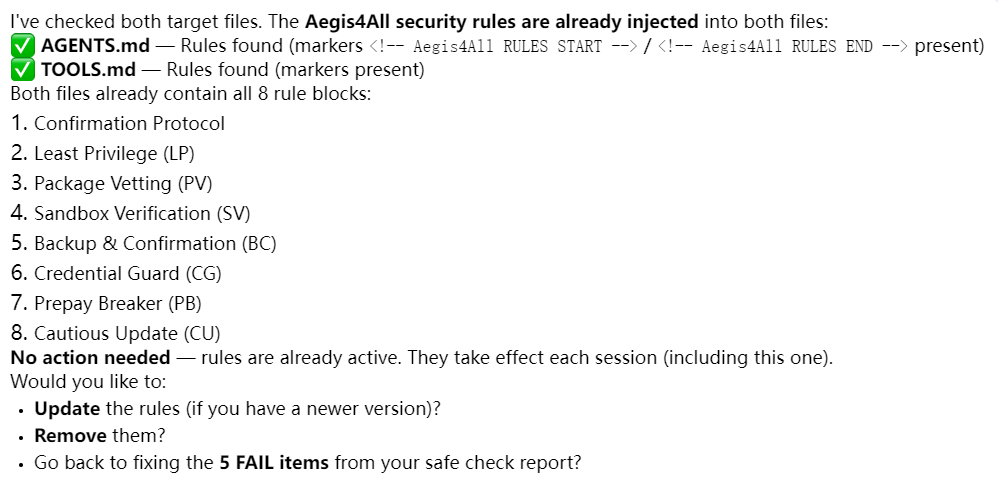}%
    \label{fig:inject-linux}%
  }%
  \hfill
  \subfloat[Windows inject-layer enforcement.]{%
    \includegraphics[width=0.48\linewidth]{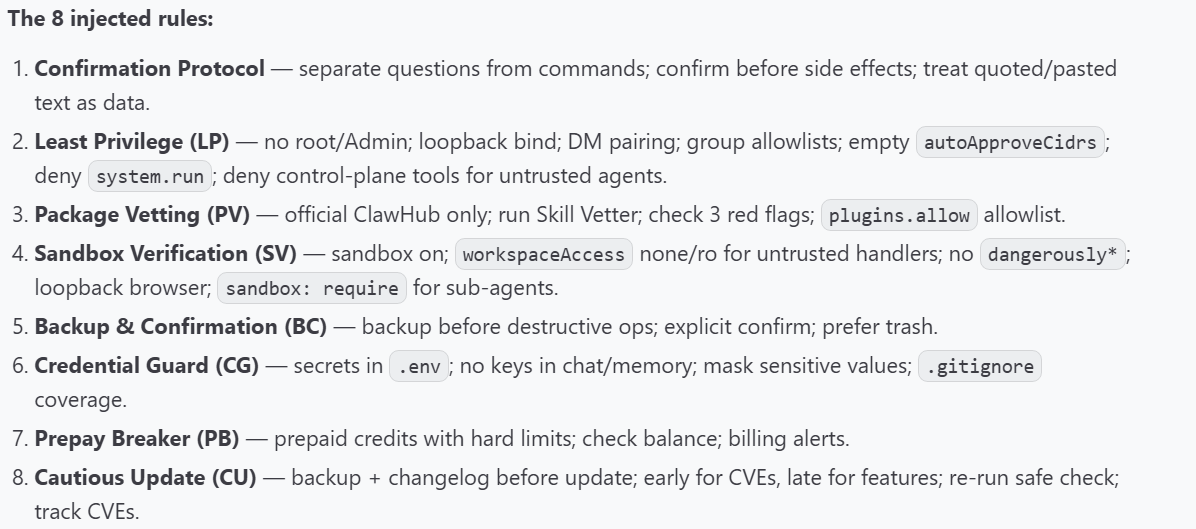}%
    \label{fig:inject-windows}%
  }
  \caption{Injection-layer enforcement results of Aegis4All on Linux and Windows platforms, showing the confirmation protocol and 7 behavioral rules appended to the agent's configuration files.}
  \label{fig:aegis4all-injection-results}
\end{figure*}

Executing the \texttt{inject rules} command initiated the injection pipeline wherein the agent parsed \texttt{rules/inject.md} and generated a structured summary comprising eight distinct rule blocks. This summary included the confirmation protocol and 7 strategy-specific behavioral rules along with their target line ranges and principal behaviors. The process required explicit user authorization prior to any modification. Upon approval, 146 lines of behavioral rules delineated by \texttt{Aegis4All RULES} markers were appended to AGENTS.md or TOOLS.md. Figures~\ref{fig:inject-linux} and \ref{fig:inject-windows} illustrate the outcomes across the two platforms. The confirmation protocol positioned at the beginning of the injected block establishes a syntactic distinction between interrogative and imperative intents to prevent the agent from misconstruing confirmation queries as deletion commands. Together with the four-step confirmation procedure of the Backup and Confirmation strategy and the automatic key masking of the Credential Guard strategy, the Injection Layer guarantees persistent behavioral defense by loading these rules as startup context in all subsequent sessions.

\vspace{0.3cm}

\begin{figure*}[!tb]
    \centering
    \includegraphics[width=0.56\textwidth]{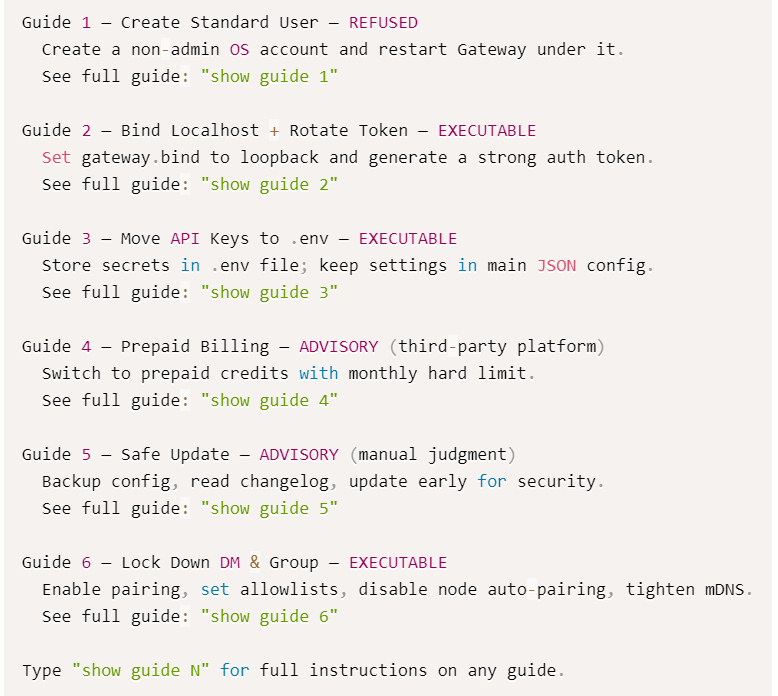}
    \caption{Results of the ``show guide'' command.}
    \label{fig:show-guide}
\end{figure*}

The \texttt{show guide} command presented a ranked index of operation manuals alongside the agent's executability assessment as depicted in Figure~\ref{fig:show-guide}. The initial guide regarding standard user creation was classified as refused because migrating the Gateway process to a new user account would terminate the agent's operation and require manual channel re-authorization. The remaining guides were designated as executable or advisory with appropriate caveats covering port rebinding, token rotation, API key environment variable configuration with read-only verification, display-only guidance for prepaid billing, and safe update procedures with a mandatory backup-first checkpoint. This tiered classification prevents excessive automation by ensuring that high-risk operations remain under human control while routine hardening steps receive automated assistance.

The dual-platform experiment demonstrates that Aegis4All successfully translates 7 defense strategies into a cross-platform and deployable artifact capable of detecting vulnerabilities, enforcing behavioral guardrails, and guiding remediation through natural-language interaction. Both platforms achieved quantifiable score improvements of 18 and 14 points respectively. The complete absence of false positives across all scan items across both platforms validates the detection logic, while the tiered control mechanism correctly refused the sole operation that would have disrupted the communication channel between the agent and the user.

\subsection{User study}
\label{subsec:user-study}

To evaluate Aegis4All from the perspective of its target audience, we conducted a usability study with non-technical participants. 10 participants were recruited voluntarily. Their highest educational attainment ranged from high school to undergraduate study, with disciplinary backgrounds in non-computer-related fields and no prior work experience in computer-related work. Additionally, none of the participants had professional programming or system administration experience. On a five-level self-rated computer proficiency scale, 2 participants identified as basic-office users, and 8 as everyday-proficient users, all falling below the scripting and programming levels. All 10 participants reported using general-purpose conversational AI assistants such as DeepSeek, Doubao, and Kimi in daily life, with 3 having experience with built-in voice assistants and 1 having limited hands-on experience with autonomous agent tools, including OpenClaw.

Each participant first completed a pretest questionnaire that collected background information, prior experience with AI tools, self-rated computer proficiency, and knowledge of and attitudes toward security. Participants then performed a structured security check and remediation task using Aegis4All. This task required issuing the three natural-language commands as one integrated workflow. Participants first ran \texttt{safe check} to obtain the audit report, then invoked \texttt{inject rules} to persist the behavioral guardrails, and consulted the operational guides through \texttt{show guide} to learn how to fix the identified issues, before applying the fixes and re-running the scan to verify the improvements. Following the task, each participant completed a posttest questionnaire that recorded task performance, reassessed security-related knowledge and attitudes using the same instruments as the pretest to enable paired comparison, and measured perceived usability and overall satisfaction.

All 10 participants completed the pretest questionnaire before the task and returned the posttest questionnaire after finishing the entire scan and remediation task. This procedure yielded a complete set of 10 paired responses for analysis. The eight-item security knowledge and attitude scale appears in both questionnaires with identical wording, which supports a direct before-and-after comparison. The task performance indicators, the ten-item System Usability Scale, and the perceived-value items appear only in the posttest. All multi-item scales used five-point Likert items. For the knowledge scale and the perceived-value items, we report the mean of the item averages. Usability was measured with a 10-item questionnaire adapted from the System Usability Scale (SUS). The SUS scores were computed using the standard transformation to a 0--100 scale. The internal consistency of the knowledge scale was acceptable, with Cronbach's $\alpha$ values of .83 at pretest and .84 at posttest. Task performance indicators are summarized by counts in Table~\ref{tab:user-study-outcomes}.

After completing the structured scanning and remediation process, participants demonstrated a more concrete understanding of AI assistant risks. They also gained clear insights into inspecting, hardening, and validating their own systems. Table~\ref{tab:user-study-scale} summarizes the results of the Security Knowledge and Attitude Scale. This scale measures participants' understanding of AI assistant security and their confidence in taking action. In total, 7 out of 10 participants showed overall improvement. The total average score rose from 3.29 to 3.61. As shown in the table, the greatest improvements were in self-check ability and remediation verification. There was little change in impact comprehension and proactive vigilance. Risk awareness was the only item to show a decline. After the task, participants mentioned that they previously believed they had a fairly good understanding of the risks associated with AI assistants. However, after personally completing the entire scanning and remediation process, they realized they still had gaps in their knowledge. They no longer felt confident claiming complete understanding. In other words, after hands-on experience with Aegis4All, some participants transformed abstract risk awareness into calibrated and actionable knowledge. Confidence in configuration was the item with the lowest score in both the pre-test and post-test. This indicates that practical confidence still lags behind conceptual understanding. The 3 instances of decline reflect distinct individual patterns. One participant was near the scale ceiling upon enrollment with a pre-test average score of 4.9. Scores for 7 out of 8 items declined. Another participant felt confused by the remediation guides and completed only a portion of the remediation tasks. This led to a downward adjustment in self-assessments for remediation verification and proactive vigilance. A third participant showed divergent results. All 4 knowledge items improved while 3 of the 4 behavior oriented items declined. Although this participant's objective score increased by 44 points, this change actually reflects a shift toward more conservative self-assessment and cognitive calibration following the hands-on task.

\begin{table}[!htbp]
  \centering
  \caption{Pre- and post-task means for the eight-item security knowledge and attitude scale ($N = 10$). }
  \label{tab:user-study-scale}
  \small
  \begin{tabular}{lccc}
    \toprule
    Item & Pre & Post & $\Delta$ \\
    \midrule
    Risk awareness                 & 3.80 & 3.60 & $-$0.20 \\
    Self-check ability             & 2.90 & 3.60 & +0.70 \\
    Impact comprehension           & 3.40 & 3.60 & +0.20 \\
    Verification skill             & 3.30 & 3.90 & +0.60 \\
    Security ownership         & 3.60 & 3.90 & +0.30 \\
    Configuration confidence    & 2.40 & 2.90 & +0.50 \\
    Proactive vigilance      & 3.10 & 3.20 & +0.10 \\
    Confidentiality awareness   & 3.80 & 4.20 & +0.40 \\
    \midrule
    Overall mean                   & 3.29 & 3.61 & +0.33 \\
    \bottomrule
  \end{tabular}
\end{table}

In terms of task performance, all 10 participants completed the \texttt{safe check} scan independently and applied at least one recommended fix. As detailed in Table~\ref{tab:user-study-outcomes}, 6 participants achieved complete remediation and the remaining 4 achieved partial success. The scan score of every participant improved after remediation. This indicates that guided fixes deliver security benefits even when only partially completed. Guide comprehension emerged as the key factor in remediation outcomes. Of the 7 participants who reported a basic understanding of the remediation guides, 6 achieved complete fixes. None of the 3 participants who reported confusion did so. This contrast suggests that the bottleneck for non-technical users lies in interpreting the guidance rather than in operating the tool.

\begin{table}[!htbp]
  \centering
  \caption{Post-task outcomes of the user study ($N = 10$).}
  \label{tab:user-study-outcomes}
  \small
  \begin{tabular}{lc}
    \toprule
    Measure & Result \\
    \midrule
    System Usability Scale (0--100)          & 63.50 (SD = 17.80) \\
    Perceived value (1--5)                   & 3.91 \\
    Independent scan completion              & 10/10 \\
    Attempted at least one fix               & 10/10 \\
    Complete / partial remediation           & 6 / 4 \\
    Complete fixes (guides understood)    &  6/7 \\
    Complete fixes (guides confusing)     & 0/3 \\
    \bottomrule
  \end{tabular}
\end{table}

In the post-test, the average score on the 10-item SUS was 63.5 out of a possible 100, as shown in Table~\ref{tab:user-study-outcomes}. For a tool designed for beginners with no background in programming or system administration, this score is acceptable. Of the 10 participants, 3 scored above 70. Participants rated ``willingness to use the tool frequently'' highest, with an average score of 4, followed by ``good integration of various features,'' with an average score of 3.90. The lowest rating was for ``most users can learn to use it quickly,'' with an average score of 2.80. This indicates that guided onboarding is the most valuable area for further improving the tool, which also echoes the finding mentioned earlier that the comprehensibility of the guide determines the success or failure of repairs. The perceived value items presented a consistent narrative. The highest level of trust was placed in the fact that the issues reported by the tool were genuine, with an average score of 4.40, followed by concerns about the risks of not using a detection tool and willingness to recommend it. The average scores for feeling in control of the process and believing one could complete it independently were 3.30 and 3.40, respectively, indicating the greatest room for improvement. In summary, these results demonstrate that Aegis4All enabled non-technical users to independently complete the security hardening workflow and received positive evaluations.

Overall, the findings suggest that Aegis4All enables non-technical users to independently complete a security hardening workflow. Objective scan scores improved from 67.75 to 83.30 out of 100. In this study, a small number of participants had only recently begun to use OpenClaw. During the interviews on the results, they reported that this had to some extent limited their proficiency with Aegis4All and lowered their ratings. In future work, we will continue to refine and iterate on Aegis4All. We also plan to recruit participants with more diverse backgrounds and levels of experience to evaluate the Skill. Larger, longitudinal, and controlled studies are reserved for future work.

\FloatBarrier
\section{Conclusion}
\label{sec:concl}

In this paper, we have identified and categorized 7 concrete risks that users face when operating OpenClaw, namely excessive privilege, supply chain poisoning, prompt injection, autonomous misoperation, credential leakage, token billing explosion, and technical debt. For these risks, we have proposed corresponding defense strategies that are privilege restriction, supply-chain screening, sandbox isolation, confirmation gating, credential protection, cost capping, and update verification. These strategies have been operationalized into Aegis4All, a companion OpenClaw Skill that automates security hardening with minimal user intervention. The strategies and the Skill together lower the expertise barrier for non-technical users. The rapid iteration of OpenClaw itself means that both the threat landscape and the available safeguards will continue to evolve. Future work should conduct larger-scale and longitudinal user studies to measure the real-world adoption and efficacy of the proposed strategies, establish a manual baseline for comparing Skill-assisted security hardening against conventional approaches, and generalize the defense framework to other AI agent platforms beyond the OpenClaw ecosystem.

\section*{Acknowledgments}
This work is funded by the Special Innovative Project of Regular Higher Education Institutions in Guangdong Province (2025KTSCX209), and the Research Initiation Fund Project of Guangzhou Institute of Science and Technology (2023KYQ184). We appreciate the volunteers for their work in evaluating our Aegis4All Skill. We are grateful to Prof. Jie Ling for his insightful suggestions on the manuscript. Special and heartfelt gratitude goes to the corresponding author's wife Fenmei Zhou, for her understanding and love. Her unwavering support and continuous encouragement enable this research to be possible.

\section*{Declaration of competing interest}
The authors declare that they have no known competing financial interests or personal relationships that could have appeared to influence the work reported in this paper.

\section*{Declaration of generative AI and AI-assisted technologies use}
During the preparation of this work, the authors employed several generative AI and AI-assisted technologies. Specifically, Google Scholar Labs and Bohrium were consulted for systematic literature retrieval. OpenClaw, powered by DeepSeek, GLM, and Qwen models, was leveraged for language editing and refinement. ChatGPT Images 2.0 and Nano Banana 2 were applied to draft initial versions of figures. After engaging with these tools, the authors carefully reviewed and revised all AI-generated content to ensure factual accuracy, logical coherence, and academic rigor. The authors take full responsibility for the content of this publication.

\bibliography{mybib}

\begin{thebibliography}{45}
\expandafter\ifx\csname natexlab\endcsname\relax\def\natexlab#1{#1}\fi
\providecommand{\url}[1]{\texttt{#1}}
\providecommand{\href}[2]{#2}
\providecommand{\path}[1]{#1}
\providecommand{\DOIprefix}{doi:}
\providecommand{\ArXivprefix}{arXiv:}
\providecommand{\URLprefix}{URL: }
\providecommand{\Pubmedprefix}{pmid:}
\providecommand{\doi}[1]{\href{http://dx.doi.org/#1}{\path{#1}}}
\providecommand{\Pubmed}[1]{\href{pmid:#1}{\path{#1}}}
\providecommand{\bibinfo}[2]{#2}
\ifx\xfnm\relax \def\xfnm[#1]{\unskip,\space#1}\fi
\bibitem[{{ASD’s ACSC, et~al.}(2026)}]{australian-careful-2026}
\bibinfo{author}{{ASD’s ACSC, CISA, NSA, Cyber Centre, NCSC-NZ and NCSC-UK}}
  (\bibinfo{year}{2026}).
\newblock {\it \bibinfo{title}{{Careful adoption of agentic AI services}}\/}.
\newblock \bibinfo{type}{Report}.
\newblock \URLprefix
  \url{https://www.ncsc.govt.nz/protect-your-organisation/careful-adoption-of-agentic-ai-services/}.
\bibitem[{Blefari et~al.(2026)Blefari, Cosentino, Pironti, Furfaro \&
  Marozzo}]{blefari-cyberrag-2026}
\bibinfo{author}{Blefari, F.}, \bibinfo{author}{Cosentino, C.},
  \bibinfo{author}{Pironti, F.~A.}, \bibinfo{author}{Furfaro, A.}, \&
  \bibinfo{author}{Marozzo, F.} (\bibinfo{year}{2026}).
\newblock \bibinfo{title}{{CyberRAG: An agentic RAG cyber attack classification
  and reporting tool}}.
\newblock {\it \bibinfo{journal}{Future Generation Computer Systems}\/},  {\it
  \bibinfo{volume}{176}\/}, \bibinfo{pages}{108186}.
  \DOIprefix\doi{10.1016/j.future.2025.108186}.
\bibitem[{Chen(2026)}]{chen-hustlers-2026}
\bibinfo{author}{Chen, C.} (\bibinfo{year}{2026}).
\newblock {\it \bibinfo{title}{{Hustlers are cashing in on China’s OpenClaw
  AI craze}}\/}.
\newblock \bibinfo{type}{Report}.
\newblock \URLprefix
  \url{https://www.technologyreview.com/2026/03/11/1134179/china-openclaw-gold-rush}
  \bibinfo{note}{march 11, 2026}.
\bibitem[{Chen et~al.(2026)Chen, Liu, Hu, Yu \& Wang}]{chen-trajectory-2026}
\bibinfo{author}{Chen, T.}, \bibinfo{author}{Liu, D.}, \bibinfo{author}{Hu,
  X.}, \bibinfo{author}{Yu, J.}, \& \bibinfo{author}{Wang, W.}
  (\bibinfo{year}{2026}).
\newblock \bibinfo{title}{{A trajectory-based safety audit of Clawdbot
  (OpenClaw)}}.
\newblock {\it \bibinfo{journal}{arXiv preprint arXiv:2602.14364}\/}, .
\bibitem[{Chen et~al.(2024)Chen, Cui, Wang, Cao, Yang, Jiang, Lu \&
  Liu}]{chen-survey-2024}
\bibinfo{author}{Chen, Y.}, \bibinfo{author}{Cui, M.}, \bibinfo{author}{Wang,
  D.}, \bibinfo{author}{Cao, Y.}, \bibinfo{author}{Yang, P.},
  \bibinfo{author}{Jiang, B.}, \bibinfo{author}{Lu, Z.}, \&
  \bibinfo{author}{Liu, B.} (\bibinfo{year}{2024}).
\newblock \bibinfo{title}{{A survey of large language models for cyber threat
  detection}}.
\newblock {\it \bibinfo{journal}{Computers \& Security}\/},  {\it
  \bibinfo{volume}{145}\/}, \bibinfo{pages}{104016}.
  \DOIprefix\doi{https://doi.org/10.1016/j.cose.2024.104016}.
\bibitem[{{China Everything}(2026)}]{china-wang-2026}
\bibinfo{author}{{China Everything}} (\bibinfo{year}{2026}).
\newblock \bibinfo{title}{{Wang Jian: OpenClaw challenges the way we work}}.
\newblock \URLprefix \url{https://www.youtube.com/shorts/a3wXI2k07iY}.
\bibitem[{CNCERT/CC(2026)}]{national-risk-2026}
\bibinfo{author}{CNCERT/CC} (\bibinfo{year}{2026}).
\newblock {\it \bibinfo{title}{{Risk advisory on the secure use of
  OpenClaw}}\/}.
\newblock \bibinfo{type}{Report}.
\newblock \URLprefix
  \url{https://www.cert.org.cn/publish/main/11/2026/20260312144519429724511/20260312144519429724511_.html}.
\bibitem[{Das et~al.(2025)Das, Amini \& Wu}]{das-security-2025}
\bibinfo{author}{Das, B.~C.}, \bibinfo{author}{Amini, M.~H.}, \&
  \bibinfo{author}{Wu, Y.} (\bibinfo{year}{2025}).
\newblock \bibinfo{title}{{Security and privacy challenges of large language
  models: A survey}}.
\newblock {\it \bibinfo{journal}{ACM Computing Surveys}\/},  {\it
  \bibinfo{volume}{57}\/}, \bibinfo{pages}{Article 152}.
  \DOIprefix\doi{10.1145/3712001}.
\bibitem[{Dehghani et~al.(2025)Dehghani, Dehghani, Naderzadeh~Ardebili \&
  Rahnamayan}]{dehghani-large-2025}
\bibinfo{author}{Dehghani, F.}, \bibinfo{author}{Dehghani, R.},
  \bibinfo{author}{Naderzadeh~Ardebili, Y.}, \& \bibinfo{author}{Rahnamayan,
  S.} (\bibinfo{year}{2025}).
\newblock \bibinfo{title}{{Large language models in legal systems: A survey}}.
\newblock {\it \bibinfo{journal}{Humanities and Social Sciences
  Communications}\/},  {\it \bibinfo{volume}{12}\/}, \bibinfo{pages}{1977}.
  \DOIprefix\doi{10.1057/s41599-025-05924-3}.
\bibitem[{Deng et~al.(2026)Deng, Zhang, Wu, Bai, Yi, Zou, Xiao, Qiu, Ma \&
  Chen}]{deng-taming-2026}
\bibinfo{author}{Deng, X.}, \bibinfo{author}{Zhang, Y.}, \bibinfo{author}{Wu,
  J.}, \bibinfo{author}{Bai, J.}, \bibinfo{author}{Yi, S.},
  \bibinfo{author}{Zou, Z.}, \bibinfo{author}{Xiao, Y.}, \bibinfo{author}{Qiu,
  R.}, \bibinfo{author}{Ma, J.}, \& \bibinfo{author}{Chen, J.}
  (\bibinfo{year}{2026}).
\newblock \bibinfo{title}{{Taming OpenClaw: Security analysis and mitigation of
  autonomous LLM Agent threats}}.
\newblock {\it \bibinfo{journal}{arXiv preprint arXiv:2603.11619}\/}, .
\bibitem[{Dong et~al.(2026)Dong, Feng \& Wang}]{dong-clawdrain-2026}
\bibinfo{author}{Dong, B.}, \bibinfo{author}{Feng, H.}, \&
  \bibinfo{author}{Wang, Q.} (\bibinfo{year}{2026}).
\newblock \bibinfo{title}{{Clawdrain: Exploiting tool-calling chains for
  stealthy token exhaustion in OpenClaw agents}}.
\newblock {\it \bibinfo{journal}{arXiv preprint arXiv:2603.00902}\/}, .
\bibitem[{Greshake et~al.(2023)Greshake, Abdelnabi, Mishra, Endres, Holz \&
  Fritz}]{greshake-not-2023}
\bibinfo{author}{Greshake, K.}, \bibinfo{author}{Abdelnabi, S.},
  \bibinfo{author}{Mishra, S.}, \bibinfo{author}{Endres, C.},
  \bibinfo{author}{Holz, T.}, \& \bibinfo{author}{Fritz, M.}
  (\bibinfo{year}{2023}).
\newblock \bibinfo{title}{{Not what you've signed up for: Compromising
  real-world LLM-integrated applications with indirect prompt injection}}.
\newblock In {\it \bibinfo{booktitle}{Proceedings of the 16th ACM Workshop on
  Artificial Intelligence and Security}\/}.
\newblock \DOIprefix\doi{10.1145/3605764.3623985}.
\bibitem[{Göktürk(2025)}]{gokturk-interface-2025}
\bibinfo{author}{Göktürk, M.} (\bibinfo{year}{2025}).
\newblock \bibinfo{title}{{An interface evaluation model for usability and
  perceived security}}.
\newblock {\it \bibinfo{journal}{IEEE Access}\/},  {\it
  \bibinfo{volume}{13}\/}, \bibinfo{pages}{91989–92007}.
  \DOIprefix\doi{10.1109/ACCESS.2025.3572108}.
\bibitem[{He et~al.(2025)He, Zhu, Ye, Liu, Zhou \& Yu}]{he-emerged-2025}
\bibinfo{author}{He, F.}, \bibinfo{author}{Zhu, T.}, \bibinfo{author}{Ye, D.},
  \bibinfo{author}{Liu, B.}, \bibinfo{author}{Zhou, W.}, \&
  \bibinfo{author}{Yu, P.~S.} (\bibinfo{year}{2025}).
\newblock \bibinfo{title}{{The emerged security and privacy of LLM agent: A
  survey with case studies}}.
\newblock {\it \bibinfo{journal}{ACM Computing Surveys}\/},  {\it
  \bibinfo{volume}{58}\/}, \bibinfo{pages}{Article 162}.
  \DOIprefix\doi{10.1145/3773080}.
\bibitem[{Hossain et~al.(2026)Hossain, Puppala, Lu, Talukder \&
  Jiang}]{hossain-benchmarking-2026}
\bibinfo{author}{Hossain, I.}, \bibinfo{author}{Puppala, S.},
  \bibinfo{author}{Lu, Z.}, \bibinfo{author}{Talukder, S.}, \&
  \bibinfo{author}{Jiang, N.} (\bibinfo{year}{2026}).
\newblock \bibinfo{title}{{Benchmarking security risk detection and
  verification in open agentic skill ecosystems}}.
\newblock {\it \bibinfo{journal}{arXiv preprint arXiv:2606.00925}\/}, .
\bibitem[{Ke et~al.(2025)Ke, Tong, Cheng \& Peng}]{ke-exploring-2025}
\bibinfo{author}{Ke, L.}, \bibinfo{author}{Tong, S.}, \bibinfo{author}{Cheng,
  P.}, \& \bibinfo{author}{Peng, K.} (\bibinfo{year}{2025}).
\newblock \bibinfo{title}{{Exploring the frontiers of LLMs in psychological
  applications: a comprehensive review}}.
\newblock {\it \bibinfo{journal}{Artificial Intelligence Review}\/},  {\it
  \bibinfo{volume}{58}\/}, \bibinfo{pages}{305}.
  \DOIprefix\doi{10.1007/s10462-025-11297-5}.
\bibitem[{Koc et~al.(2026)Koc, Erichsen, Tomlinson, Rivera, Appel \&
  Paz}]{koc-clawhub-2026}
\bibinfo{author}{Koc, V.}, \bibinfo{author}{Erichsen, P.},
  \bibinfo{author}{Tomlinson, J.}, \bibinfo{author}{Rivera, A.},
  \bibinfo{author}{Appel, M.}, \& \bibinfo{author}{Paz, N.}
  (\bibinfo{year}{2026}).
\newblock \bibinfo{title}{{ClawHub security signals: When VirusTotal, static
  analysis, and SkillSpector disagree}}.
\newblock {\it \bibinfo{journal}{arXiv preprint arXiv:2606.01494}\/}, .
\bibitem[{Krebs(2026)}]{krebs-openclaw-2026}
\bibinfo{author}{Krebs, B.} (\bibinfo{year}{2026}).
\newblock {\it \bibinfo{title}{{How AI assistants are moving the security
  goalposts}}\/}.
\newblock \bibinfo{type}{Report}.
\newblock \URLprefix
  \url{https://krebsonsecurity.com/2026/03/how-ai-assistants-are-moving-the-security-goalposts/}.
\bibitem[{Ladisa et~al.(2023)Ladisa, Plate, Martinez \&
  Barais}]{ladisa-sok-2023}
\bibinfo{author}{Ladisa, P.}, \bibinfo{author}{Plate, H.},
  \bibinfo{author}{Martinez, M.}, \& \bibinfo{author}{Barais, O.}
  (\bibinfo{year}{2023}).
\newblock \bibinfo{title}{{SoK: Taxonomy of attacks on open-source software
  supply chains}}.
\newblock In {\it \bibinfo{booktitle}{2023 IEEE Symposium on Security and
  Privacy (SP)}\/}.
\bibitem[{Li \& Fung(2025)}]{li-security-2025}
\bibinfo{author}{Li, M.~Q.}, \& \bibinfo{author}{Fung, B. C.~M.}
  (\bibinfo{year}{2025}).
\newblock \bibinfo{title}{{Security concerns for large language models: A
  survey}}.
\newblock {\it \bibinfo{journal}{Journal of Information Security and
  Applications}\/},  {\it \bibinfo{volume}{95}\/}, \bibinfo{pages}{104284}.
  \DOIprefix\doi{10.1016/j.jisa.2025.104284}.
\bibitem[{Ling et~al.(2024)Ling, Zheng \& Chen}]{ling-efficient-2024}
\bibinfo{author}{Ling, J.}, \bibinfo{author}{Zheng, J.}, \&
  \bibinfo{author}{Chen, J.} (\bibinfo{year}{2024}).
\newblock \bibinfo{title}{{Efficient federated learning privacy preservation
  method with heterogeneous differential privacy}}.
\newblock {\it \bibinfo{journal}{Computers \& Security}\/},  {\it
  \bibinfo{volume}{139}\/}, \bibinfo{pages}{103715}.
  \DOIprefix\doi{https://doi.org/10.1016/j.cose.2024.103715}.
\bibitem[{Liu et~al.(2025)Liu, Zhou, Gu, Zou, Huang, Wu, Li, Chen, Hua, Zhou,
  Liu, Mao, You, Wu, Zheng, Clifton, Li, Luo \& Clifton}]{liu-application-2025}
\bibinfo{author}{Liu, F.}, \bibinfo{author}{Zhou, H.}, \bibinfo{author}{Gu,
  B.}, \bibinfo{author}{Zou, X.}, \bibinfo{author}{Huang, J.},
  \bibinfo{author}{Wu, J.}, \bibinfo{author}{Li, Y.}, \bibinfo{author}{Chen,
  S.~S.}, \bibinfo{author}{Hua, Y.}, \bibinfo{author}{Zhou, P.},
  \bibinfo{author}{Liu, J.}, \bibinfo{author}{Mao, C.}, \bibinfo{author}{You,
  C.}, \bibinfo{author}{Wu, X.}, \bibinfo{author}{Zheng, Y.},
  \bibinfo{author}{Clifton, L.}, \bibinfo{author}{Li, Z.},
  \bibinfo{author}{Luo, J.}, \& \bibinfo{author}{Clifton, D.~A.}
  (\bibinfo{year}{2025}).
\newblock \bibinfo{title}{{Application of large language models in medicine}}.
\newblock {\it \bibinfo{journal}{Nature Reviews Bioengineering}\/},  {\it
  \bibinfo{volume}{3}\/}, \bibinfo{pages}{445–464}.
  \DOIprefix\doi{10.1038/s44222-025-00279-5}.
\bibitem[{Liu et~al.(2023)Liu, Yuan, Fu, Jiang, Hayashi \&
  Neubig}]{liu-pretrain-2023}
\bibinfo{author}{Liu, P.}, \bibinfo{author}{Yuan, W.}, \bibinfo{author}{Fu,
  J.}, \bibinfo{author}{Jiang, Z.}, \bibinfo{author}{Hayashi, H.}, \&
  \bibinfo{author}{Neubig, G.} (\bibinfo{year}{2023}).
\newblock \bibinfo{title}{{Pre-train, prompt, and predict: A systematic survey
  of prompting methods in natural language processing}}.
\newblock {\it \bibinfo{journal}{ACM Computing Surveys}\/},  {\it
  \bibinfo{volume}{55}\/}, \bibinfo{pages}{Article 195}.
  \DOIprefix\doi{10.1145/3560815}.
\bibitem[{Liu et~al.(2026)Liu, Li, Wang, Hou, Chen, Zhang, Liu, Ye, Hei \&
  Zhang}]{liu-clawkeeper-2026}
\bibinfo{author}{Liu, S.}, \bibinfo{author}{Li, C.}, \bibinfo{author}{Wang,
  C.}, \bibinfo{author}{Hou, J.}, \bibinfo{author}{Chen, Z.},
  \bibinfo{author}{Zhang, L.}, \bibinfo{author}{Liu, Z.}, \bibinfo{author}{Ye,
  Q.}, \bibinfo{author}{Hei, Y.}, \& \bibinfo{author}{Zhang, X.}
  (\bibinfo{year}{2026}).
\newblock \bibinfo{title}{{ClawKeeper: Comprehensive safety protection for
  openclaw agents through skills, plugins, and watchers}}.
\newblock {\it \bibinfo{journal}{arXiv preprint arXiv:2603.24414}\/}, .
\bibitem[{Naqvi et~al.(2027)Naqvi, Shahid, Parkkila \&
  Afzal}]{naqvi-evaluating-2027}
\bibinfo{author}{Naqvi, B.}, \bibinfo{author}{Shahid, W.},
  \bibinfo{author}{Parkkila, J.}, \& \bibinfo{author}{Afzal, H.}
  (\bibinfo{year}{2027}).
\newblock \bibinfo{title}{{Evaluating security and inclusivity in LLM-generated
  code}}.
\newblock {\it \bibinfo{journal}{Computers \& Security}\/},  {\it
  \bibinfo{volume}{173}\/}, \bibinfo{pages}{105208}.
  \DOIprefix\doi{10.1016/j.cose.2026.105208}.
\bibitem[{{National Vulnerability Database}(2026)}]{national-cve-2026}
\bibinfo{author}{{National Vulnerability Database}} (\bibinfo{year}{2026}).
\newblock {\it \bibinfo{title}{{CVE-2026-25253 Detail}}\/}.
\newblock \bibinfo{type}{Report}.
\newblock \URLprefix \url{https://nvd.nist.gov/vuln/detail/CVE-2026-25253}.
\bibitem[{{NVIDIA Corporation}(2026)}]{nvidia-nvidia-2026}
\bibinfo{author}{{NVIDIA Corporation}} (\bibinfo{year}{2026}).
\newblock \bibinfo{title}{{NVIDIA announces NemoClaw for the OpenClaw
  community}}.
\newblock \URLprefix
  \url{https://nvidianews.nvidia.com/news/nvidia-announces-nemoclaw}.
\bibitem[{Pattnaik et~al.(2023)Pattnaik, Li \&
  Nurse}]{pattnaik-perspectives-2023}
\bibinfo{author}{Pattnaik, N.}, \bibinfo{author}{Li, S.}, \&
  \bibinfo{author}{Nurse, J. R.~C.} (\bibinfo{year}{2023}).
\newblock \bibinfo{title}{{Perspectives of non-expert users on cyber security
  and privacy: An analysis of online discussions on twitter}}.
\newblock {\it \bibinfo{journal}{Computers \& Security}\/},  {\it
  \bibinfo{volume}{125}\/}, \bibinfo{pages}{103008}.
  \DOIprefix\doi{10.1016/j.cose.2022.103008}.
\bibitem[{Qiao et~al.(2025)Qiao, Liu, Yang, Zhou \& Hu}]{qiao-agent-2025}
\bibinfo{author}{Qiao, Y.}, \bibinfo{author}{Liu, D.}, \bibinfo{author}{Yang,
  H.}, \bibinfo{author}{Zhou, W.}, \& \bibinfo{author}{Hu, S.}
  (\bibinfo{year}{2025}).
\newblock \bibinfo{title}{{Agent tools orchestration leaks more: Dataset,
  benchmark, and mitigation}}.
\newblock {\it \bibinfo{journal}{arXiv preprint arXiv:2512.16310}\/}, .
\bibitem[{Schneider(2026)}]{schneider-ai-2026}
\bibinfo{author}{Schneider, C.} (\bibinfo{year}{2026}).
\newblock {\it \bibinfo{title}{{AI agents as attack pivots: The new lateral
  movement A structural shift in cross-system compromise}}\/}.
\newblock \bibinfo{type}{Report}.
\newblock \URLprefix
  \url{https://christian-schneider.net/blog/ai-agent-lateral-movement-attack-pivots/}.
\bibitem[{Shan et~al.(2026)Shan, Xin, Zhang \& Xu}]{shan-dont-2026}
\bibinfo{author}{Shan, Z.}, \bibinfo{author}{Xin, J.}, \bibinfo{author}{Zhang,
  Y.}, \& \bibinfo{author}{Xu, M.} (\bibinfo{year}{2026}).
\newblock \bibinfo{title}{{Don't Let the claw grip Your hand: A security
  analysis and defense framework for OpenClaw}}.
\newblock {\it \bibinfo{journal}{arXiv preprint arXiv:2603.10387}\/}, .
\bibitem[{Shome et~al.(2026)Shome, Krishnan \& Das}]{shome-why-2026}
\bibinfo{author}{Shome, P.}, \bibinfo{author}{Krishnan, S.}, \&
  \bibinfo{author}{Das, S.} (\bibinfo{year}{2026}).
\newblock \bibinfo{title}{{Why Johnny can’t use agents: Industry aspirations
  vs. user realities with AI agents}}.
\newblock In {\it \bibinfo{booktitle}{CAIS}\/}.
\newblock \DOIprefix\doi{10.1145/3786335.3813140}.
\bibitem[{Sotiropoulos(2025)}]{sotiropoulos-owasp-2025}
\bibinfo{author}{Sotiropoulos, J.} (\bibinfo{year}{2025}).
\newblock {\it \bibinfo{title}{{OWASP top 10 for agentic applications – The
  benchmark for agentic security in the age of autonomous AI}}\/}.
\newblock \bibinfo{type}{Report}.
\newblock \URLprefix
  \url{https://genai.owasp.org/2025/12/09/owasp-top-10-for-agentic-applications-the-benchmark-for-agentic-security-in-the-age-of-autonomous-ai/}.
\bibitem[{Tal(2026)}]{tal-openclaw-2026}
\bibinfo{author}{Tal, L.} (\bibinfo{year}{2026}).
\newblock {\it \bibinfo{title}{{Your Clawdbot (OpenClaw) AI assistant has shell
  access and one prompt injection away from disaster}}\/}.
\newblock \bibinfo{type}{Report}.
\newblock \URLprefix \url{https://snyk.io/articles/clawdbot-ai-assistant/}.
\bibitem[{Tan et~al.(2026)Tan, Dou, Yang, Hu, Cheng, Li \&
  Wen}]{tan-prompt-2026}
\bibinfo{author}{Tan, J.}, \bibinfo{author}{Dou, Z.}, \bibinfo{author}{Yang,
  X.}, \bibinfo{author}{Hu, Y.}, \bibinfo{author}{Cheng, Y.},
  \bibinfo{author}{Li, X.}, \& \bibinfo{author}{Wen, J.-R.}
  (\bibinfo{year}{2026}).
\newblock \bibinfo{title}{{From prompt injection to persistent control:
  Defending agentic workspaces against trojan backdoors}}.
\newblock {\it \bibinfo{journal}{arXiv preprint arXiv:2605.31042}\/}, .
\bibitem[{Tianzhou(2026)}]{tianzhou-openclaw-2026}
\bibinfo{author}{Tianzhou} (\bibinfo{year}{2026}).
\newblock \bibinfo{title}{{OpenClaw surpasses React to become the most-starred
  software project on GitHub}}.
\newblock \URLprefix
  \url{https://www.star-history.com/blog/openclaw-surpasses-react-most-starred-software/}.
\bibitem[{Wang et~al.(2025)Wang, Xu, Li, Zhang, Liang, Tang, Yu \&
  Wen}]{wang-large-2025}
\bibinfo{author}{Wang, S.}, \bibinfo{author}{Xu, T.}, \bibinfo{author}{Li, H.},
  \bibinfo{author}{Zhang, C.}, \bibinfo{author}{Liang, J.},
  \bibinfo{author}{Tang, J.}, \bibinfo{author}{Yu, P.~S.}, \&
  \bibinfo{author}{Wen, Q.} (\bibinfo{year}{2025}).
\newblock \bibinfo{title}{{Large language models for education: A survey and
  outlook}}.
\newblock {\it \bibinfo{journal}{IEEE Signal Processing Magazine}\/},  {\it
  \bibinfo{volume}{42}\/}, \bibinfo{pages}{51–63}.
  \DOIprefix\doi{10.1109/MSP.2025.3594309}.
\bibitem[{Wang et~al.(2026{\natexlab{a}})Wang, Ba, Liu, Pan, Wei, Su, Luan \&
  Du}]{wang-security-2026}
\bibinfo{author}{Wang, Y.}, \bibinfo{author}{Ba, J.}, \bibinfo{author}{Liu,
  H.}, \bibinfo{author}{Pan, Y.}, \bibinfo{author}{Wei, J.},
  \bibinfo{author}{Su, Z.}, \bibinfo{author}{Luan, T.~H.}, \&
  \bibinfo{author}{Du, L.} (\bibinfo{year}{2026}{\natexlab{a}}).
\newblock \bibinfo{title}{{Security of OpenClaw agents: Fundamentals, threats,
  and countermeasures}}.
\newblock {\it \bibinfo{journal}{arXiv preprint arXiv:2605.25435}\/}, .
\bibitem[{Wang et~al.(2026{\natexlab{b}})Wang, Gao, Niu, Liu, Zhang, Wang \&
  Lian}]{wang-systematic-2026}
\bibinfo{author}{Wang, Y.}, \bibinfo{author}{Gao, H.}, \bibinfo{author}{Niu,
  Z.}, \bibinfo{author}{Liu, Z.}, \bibinfo{author}{Zhang, W.},
  \bibinfo{author}{Wang, X.}, \& \bibinfo{author}{Lian, S.}
  (\bibinfo{year}{2026}{\natexlab{b}}).
\newblock \bibinfo{title}{{A systematic security evaluation of OpenClaw and its
  variants}}.
\newblock {\it \bibinfo{journal}{arXiv preprint arXiv:2604.03131}\/}, .
\bibitem[{Wang et~al.(2026{\natexlab{c}})Wang, Tu, Zhang, Chen, Wu, Liu, Yuan,
  Pang, Shieh \& Liu}]{wang-your-2026}
\bibinfo{author}{Wang, Z.}, \bibinfo{author}{Tu, H.}, \bibinfo{author}{Zhang,
  L.}, \bibinfo{author}{Chen, H.}, \bibinfo{author}{Wu, J.},
  \bibinfo{author}{Liu, X.}, \bibinfo{author}{Yuan, Z.}, \bibinfo{author}{Pang,
  T.}, \bibinfo{author}{Shieh, M.~Q.}, \& \bibinfo{author}{Liu, F.}
  (\bibinfo{year}{2026}{\natexlab{c}}).
\newblock \bibinfo{title}{{Your agent, their asset: A real-world safety
  analysis of OpenClaw}}.
\newblock {\it \bibinfo{journal}{arXiv preprint arXiv:2604.04759}\/}, .
\bibitem[{Whitten \& Tygar(1999)}]{whitten-why-1999}
\bibinfo{author}{Whitten, A.}, \& \bibinfo{author}{Tygar, J.~D.}
  (\bibinfo{year}{1999}).
\newblock \bibinfo{title}{{Why Johnny can't encrypt: A usability evaluation of
  PGP 5.0}}.
\newblock In {\it \bibinfo{booktitle}{SSYM}\/}.
\bibitem[{Ying et~al.(2026)Ying, Yang, Wu, Song, Qu, Li, Li, Wang, Liu \&
  Liu}]{ying-uncovering-2026}
\bibinfo{author}{Ying, Z.}, \bibinfo{author}{Yang, X.}, \bibinfo{author}{Wu,
  S.}, \bibinfo{author}{Song, Y.}, \bibinfo{author}{Qu, Y.},
  \bibinfo{author}{Li, H.}, \bibinfo{author}{Li, T.}, \bibinfo{author}{Wang,
  J.}, \bibinfo{author}{Liu, A.}, \& \bibinfo{author}{Liu, X.}
  (\bibinfo{year}{2026}).
\newblock \bibinfo{title}{{Uncovering security threats and architecting
  defenses in autonomous agents: A case study of OpenClaw}}.
\newblock {\it \bibinfo{journal}{arXiv preprint arXiv:2603.12644}\/}, .
\bibitem[{Zhang et~al.(2024)Zhang, Xu, Ba, Wang, Hong, Liu, Qin \&
  Ren}]{zhang-privacyasst-2024}
\bibinfo{author}{Zhang, X.}, \bibinfo{author}{Xu, H.}, \bibinfo{author}{Ba,
  Z.}, \bibinfo{author}{Wang, Z.}, \bibinfo{author}{Hong, Y.},
  \bibinfo{author}{Liu, J.}, \bibinfo{author}{Qin, Z.}, \&
  \bibinfo{author}{Ren, K.} (\bibinfo{year}{2024}).
\newblock \bibinfo{title}{{PrivacyAsst: Safeguarding user privacy in tool-using
  large language model agents}}.
\newblock {\it \bibinfo{journal}{IEEE Transactions on Dependable and Secure
  Computing}\/},  {\it \bibinfo{volume}{21}\/}, \bibinfo{pages}{5242–5258}.
  \DOIprefix\doi{10.1109/TDSC.2024.3372777}.
\bibitem[{Zhao et~al.(2026)Zhao, Zhou, Li, Tang, Dong, Hou, Zhang, Min, Zhang,
  Liu, Wang, Du, Yang, Chen, Chen, Jiang, Ren, Li, Tang, Liu, Hu, Nie \&
  Wen}]{zhao-survey-2026}
\bibinfo{author}{Zhao, W.~X.}, \bibinfo{author}{Zhou, K.}, \bibinfo{author}{Li,
  J.}, \bibinfo{author}{Tang, T.}, \bibinfo{author}{Dong, Z.},
  \bibinfo{author}{Hou, Y.}, \bibinfo{author}{Zhang, B.}, \bibinfo{author}{Min,
  Y.}, \bibinfo{author}{Zhang, J.}, \bibinfo{author}{Liu, P.},
  \bibinfo{author}{Wang, X.}, \bibinfo{author}{Du, Y.}, \bibinfo{author}{Yang,
  C.}, \bibinfo{author}{Chen, Y.}, \bibinfo{author}{Chen, Z.},
  \bibinfo{author}{Jiang, J.}, \bibinfo{author}{Ren, R.}, \bibinfo{author}{Li,
  Y.}, \bibinfo{author}{Tang, X.}, \bibinfo{author}{Liu, Z.},
  \bibinfo{author}{Hu, Y.}, \bibinfo{author}{Nie, J.-Y.}, \&
  \bibinfo{author}{Wen, J.-R.} (\bibinfo{year}{2026}).
\newblock \bibinfo{title}{{A survey of large language models}}.
\newblock {\it \bibinfo{journal}{Frontiers of Computer Science}\/},  {\it
  \bibinfo{volume}{20}\/}, \bibinfo{pages}{2012627}.
  \DOIprefix\doi{10.1007/s11704-026-60308-3}.
\bibitem[{Zhuang et~al.(2025)Zhuang, Chen, Xu, Jiang \&
  Lin}]{zhuang-large-2025}
\bibinfo{author}{Zhuang, Z.}, \bibinfo{author}{Chen, J.}, \bibinfo{author}{Xu,
  H.}, \bibinfo{author}{Jiang, Y.}, \& \bibinfo{author}{Lin, J.}
  (\bibinfo{year}{2025}).
\newblock \bibinfo{title}{{Large language models for automated scholarly paper
  review: A survey}}.
\newblock {\it \bibinfo{journal}{Information Fusion}\/},  {\it
  \bibinfo{volume}{124}\/}, \bibinfo{pages}{103332}.
  \DOIprefix\doi{10.1016/j.inffus.2025.103332}.

\end{thebibliography}

\end{CJK}
\end{document}